 \documentclass[twocolumn]{aastex631}
\shorttitle{Gaia22alz}
\shortauthors{Kato and Hachisu}

\begin{document}

% metamorphosed  metamorphosed?   and Gaia22alz

\title{Metamorphic novae. I. Gaia22alz
-- Violent relaxation starting from static evolution
}


\author[0000-0002-8522-8033]{Mariko Kato}
\affil{Department of Astronomy, Keio University,
Hiyoshi, Kouhoku-ku, Yokohama 223-8521, Japan}
\email{mariko.kato@hc.st.keio.ac.jp}

\author[0000-0002-0884-7404]{Izumi Hachisu}
\affil{Department of Earth Science and Astronomy,
College of Arts and Sciences, The University of Tokyo,
3-8-1 Komaba, Meguro-ku, Tokyo 153-8902, Japan}
%\email{izumi.hachisu@outlook.jp}

%\author{Hideyuki Saio}
%\affil{Astronomical Institute, Graduate School of Science,
% Tohoku University, Sendai 980-8578, Japan}
% \email{saio@astr.tohoku.ac.jp}

%\and


%\altaffiltext{1}{Visiting Astronomer, Cerro Tololo Inter-American Observatory\
.
%CTIO is operated by AURA, Inc.\ under contract to the National Science
%\altaffiltext{2}{Society of Fellows, Harvard University.}
%\altaffiltext{3}{present address: Center for Astrophysics,
%    60 Garden Street, Cambridge, MA 02138}
%\altaffiltext{4}{Visiting Programmer, Space Telescope Science Institute}
%\altaffiltext{5}{Patron, Alonso's Bar and Grill}


\begin{abstract}
Gaia22alz (nova Vel 2022) 
shows characteristic properties in its light curves and spectral evolution  
similar to slow novae HR Del, V723 Cas, and V5558 Sgr. 
Their outburst begins in a way of hydrostatic expansion 
with no mass loss, followed by large amplitude 
($\Delta V \sim  2$ mag) multiple peaks around the mean magnitude 
of $M_V \sim -5.5$, and eventually it decays similarly 
to normal classical novae. 
We present a light curve model of Gaia22alz 
based on our metamorphic nova theory: a nova starts from a static
expansion and then metamorphoses to an optically thick wind evolution
through a large amplitude multiple peak phase
as a violent relaxation process.
Our theoretical light curve for the rising phase consistently explains 
the slow rise and early spectral evolution of Gaia22alz. 
Gaia22als is a close binary as well as other three novae, 
which is consistent with our metamorphic nova theory 
that the companion motion in the nova envelope
triggers the metamorphic transition. 
Our results do not support E. Aydi et al.'s explanation that the slow 
optical rise of Gaia22alz is owing to the irradiation of X-ray/UV flash.
\end{abstract}
% up to 16 lines

%% Keywords should appear after the \enda{abstract} command. The uncommented



\section{Introduction} \label{sec_introduction}

% Fig.1
\begin{figure*}
%\epsscale{1.05}
\epsscale{0.9}
%\plotone{light.compari.2026.ps}
\plotone{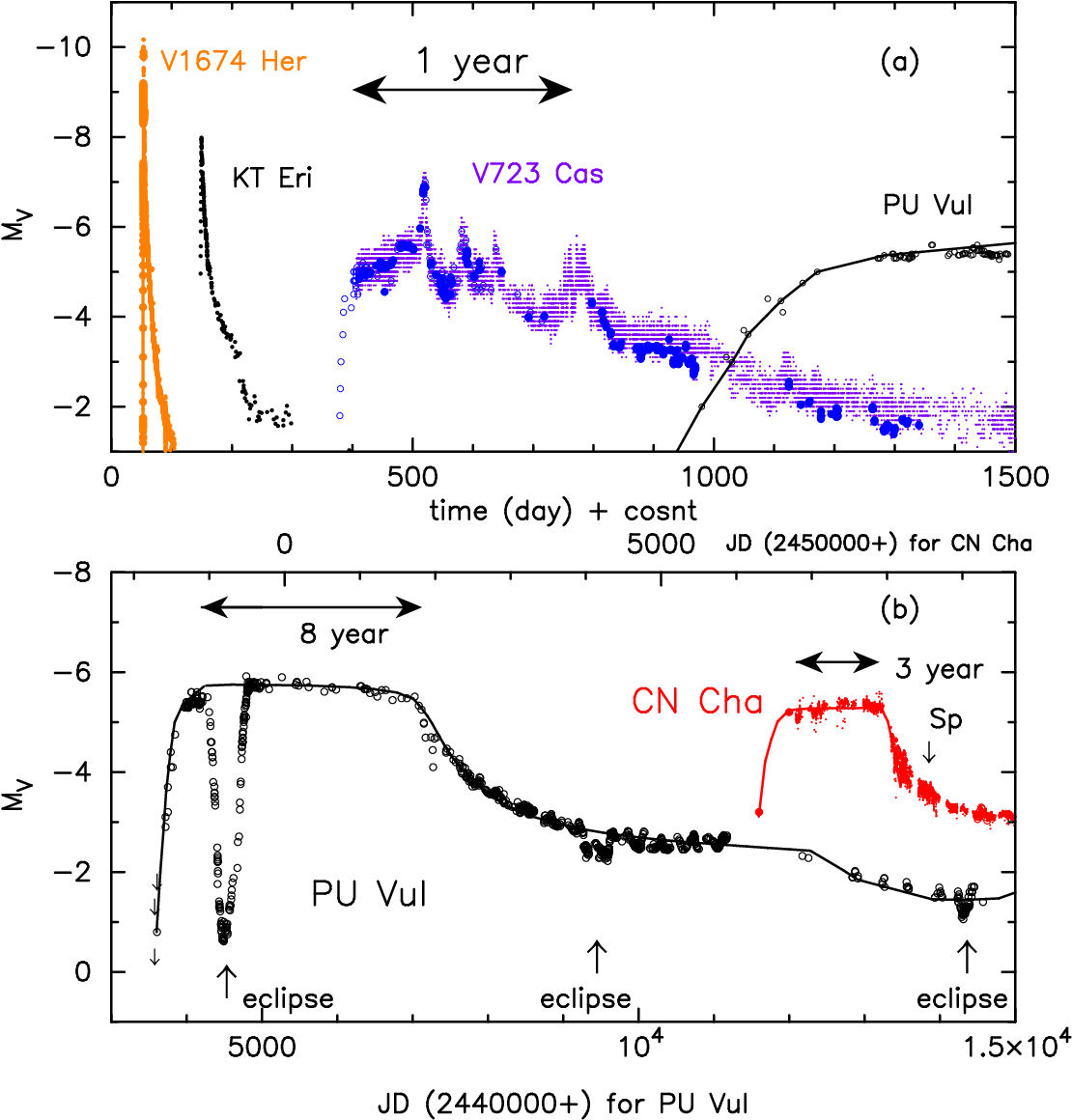}
\caption{
(a) Comparison of $V$ light curves among four novae in the 
absolute $V$ magnitude (see Table \ref{table_novae} for the distance
 modulus in the $V$ band of each nova). 
V1674 Her: the optical data are taken from \citet{qui24} and the archive
of the American Association of Variable Star Observers (AAVSO).  The 
orange line indicates the model light curve for a 1.35 $M_\sun$ WD 
with the mass accretion rate to the WD of 
$\dot M_{\rm acc}= 1\times 10^{-11}~M_\sun$~yr$^{-1}$ \citep{kat25v1674her}.
KT Eri: the optical data are the same as those in Figure 1 of
\citet{hac25kw}.  V723 Cas: filled circles are taken from \citet{cho97},
small open circles and purple dots are the same as those in Figure 1 of
\citet{kat23}.
%%%(IAUC Nos.
%%%6213, 6214, 6227, 6233, 6256, 6283, 6331, 6358, 6428), and 
%%%purple dots ($V$ and visual; AAVSO).
%The black line is a nova outburst model for a 0.6 $M_\sun$ WD
%with the solar composition envelope \citep{kat11drag}. 
 PU Vul: the optical data and 0.6 $M_\sun$ WD model 
are the same as those in Figure 6 of \citet{kat11}. 
(b) PU Vul: the optical data and 0.6 $M_\sun$ WD model 
are the same as those in Figure 6 of \citet{kat11} and 
Figure 8 of \citet{kat12}.  
CN Cha: the optical data taken from \citet{lan20} and AAVSO. 
The red line denotes the 0.57 $M_\sun$ WD for CN Cha taken from 
Figure 2 of \citet{kat23}.
}\label{novae}
% source:   nova/cncha/light.compari.2026.wip
\end{figure*}

% Table.1
%\startlongtable
\begin{deluxetable*}{lllllllll}
%\begin{deluxetable*}{llllllllllcclllll}
\tabletypesize{\scriptsize}
\tablecaption{Nova Properties
\label{table_novae}}
%\tablewidth{0pt}
\tablehead{
\colhead{Object} &
\colhead{Outburst}&
\colhead{$(m-M)_V$}&
\colhead{$E(B-V)$}&
\colhead{$d$} &
\colhead{$M_{\rm WD}$} &
\colhead{binary} &
\colhead{$P_{\rm orb}$} &
\colhead{reference for $(m-M)_V$} \\
%%%%%\colhead{reference\tablenotemark{a}} \\
\colhead{ }&
\colhead{ year } &
\colhead{ }&
\colhead{ }&
\colhead{(kpc)   } &
\colhead{($M_\sun$) }&
\colhead{($M_\sun$) }&
\colhead{(day)}& 
\colhead{ }
}
\startdata
%     &   yr  &  m-M)v  &  E(B-V)&   distance &  MWD  binary Porb  ref
V1674 Her & 2021 & 16.3 & 0.5    & 8.9  & 1.35 & close & 0.153 & \citet{kat25v1674her} \\
KT Eri    & 2009 & 13.4 & 0.08   & 4.2  & 1.3 & close & 2.62 & \citet{hac25kw} \\
HR Del    & 1967 & 10.4 & 0.15   & 0.97 & $\sim 0.6$ & close & 0.214 & \citet{hac15k} \\
V723 Cas & 1995 & 14.0  & 0.35   & 3.85 & $\sim 0.6$ & close & 0.693 & \citet{hac15k} \\
V5558 Sgr & 2007 & 13.9 &  0.7 & 2.2  & $\sim 0.6$ & close & 0.186 & \citet{hac16k} \\
% V1405 Cas & 2021 & 12.77 & 0.51 & 1.73  &0.6 &close &\\
Gaia22alz   & 2022 & 17.35 & 0.9 & 8.15  & $\sim 0.6$ & close & 1.437 & this work\\
PU Vul    & 1979 & 14.3 & 0.3 & 4.7  & $ \sim 0.6$ & wide & 4894 & \citet{kat11,kat12} \\
CN Cha    & 2012-2013 & 12.9 & 0.17  & 3.05 & $\sim 0.6$ & wide & 253 & \citet{kat23} 
\enddata
%\tablenotetext{a}{Reference for the distance modulus in the $V$ band,
%1: \citet{kat25v1674her}, 
%2: \citet{hac25kw}, 
%3: \citet{hac15k}, 
%4: \citet{hac16k}, 
%5: \citet{kat11,kat12}, 
%6: \citet{kat23}.
%}
%%%\tablenotetext{b}{ TESSExtractor \citep{ser26}, see also Section \ref{sec_companion}
%}
%\tablenotetext{c}{ a dwarf companion is suggested by \citet{ayd23}}
%\tablenotetext{d}{Increased carbon mass-fraction by 0.1 at ignition.}
\end{deluxetable*}

%  Porb (day) =  1.437 day  34.488 hour
%  M1   M2      a (Ro)      R2*/a     R2(sol)      R1(sol)
%  0.6  0.2  4.97361982  0.288981148  1.43728237  2.36776691
% log T=  4.485 log L/Lo=  3.20735736 R=  1.00034853E+11 (cm)  1.43728237 (Ro)
%  3.14159265  3.46163939E+11  124156.8
% Vorb=  17518268.6  7.24349118

%%% Gaia DR3 parallax and distance %%%
% HR Del 0.15     1813953083546374144   1.1157647 0.021395849     12.284572   867.699  851.174 886.234  872.450 858.251 886.891 10022

% V5558 Sgr  0.05 4095547845604621824  0.46682113 0.028908474     14.342856   1983.248   1874.266  2112.878  1964.377  1843.253  2075.859 10033

% V723 Cas  0.06  411250132279713280  0.18332711  0.035208672     15.65771    4498.632  3890.780  5385.345    4186.143  3655.777  4733.649 10033

% PU Vul  0.11    1829139027066224640  0.19088939 0.03931089      11.131271   4233.942  3684.844  4840.693    4090.013  3491.823  4629.874 10033

% CN Cha  0.14    5199012973569013120  0.3222786  0.019822778      7.44369    3051.338  2876.670  3245.439    3062.400  2854.303  3293.817 10033

% Fig.2
\begin{figure*}
%\epsscale{1.15}
\epsscale{0.75}
%\plotone{light.4nova.f.ps}
\plotone{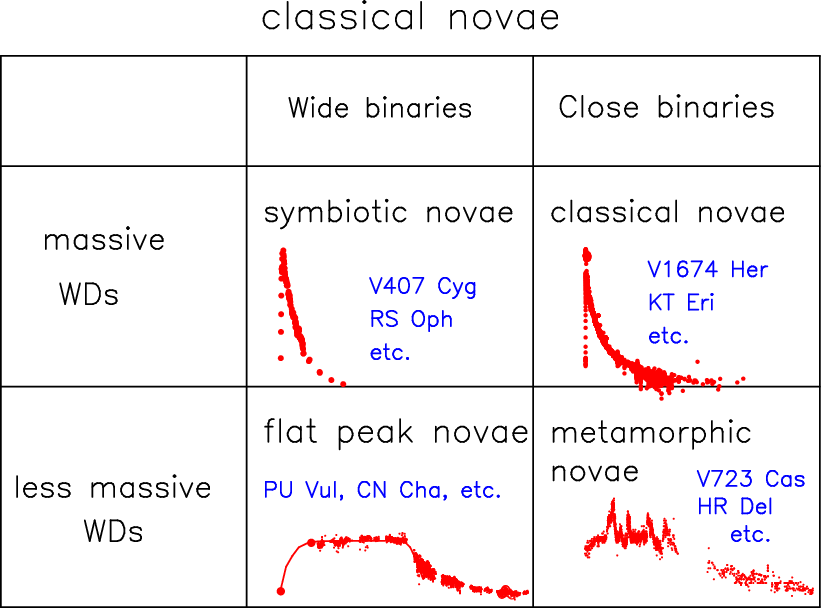}
\caption{ 
Observational aspects of metamorphic novae.  We categorize the metamorphic
novae among other classes of novae, based on the theoretical interpretation
of metamorphic transition \citep{kat11}. 
A metamorphic nova occurs on a low mass WD in a close binary.
Here, classical novae involve all the hydrogen shell-flash phenomena on a WD. 
%Category of our nova classification based on their light curve shapes,
% binary natures, and WD masses.  Classical novae show a rapid rise,
%fast massive wind ($v_{\rm wind}\gtrsim 300$ km 
%s$^{-1}$), and relatively fast decline. 
In massive WDs, a strong optically-thick wind mass-loss \citep{kat94h} 
occurs and makes a rapid optical decline independently of the wide
or close binary type.  Some less massive WDs show a long lasted
flat peak with  
no indication of massive winds (no P Cygni profiles) in the rising 
phase followed by a very long (longer than a few years) flat peak.
Metamorphic novae show a hybrid evolution of flat peak novae and
classical novae on a massive WD, that is, start a hydrostatic
evolution (no indication of massive winds)
in the rising phase and then metamorphose to a wind evolution after
a violent multi-peak episode.  
}\label{category}
\end{figure*}

A nova is a thermonuclear runaway event that occurs on an accreting 
white dwarf (WD) in a binary 
\citep[e.g.,][]{nar80, ibe82, pri95, sio79, spa78,taj15sn}. 
After the thermonuclear runaway sets in, the hydrogen-rich envelope on 
the WD becomes bright and expands to a giant star size. 
The WD envelope blows massive winds. 

Novae show a rich variety of their decay timescales. 
Figure \ref{novae}(a) shows four nova light curves in the order of
nova speed class,\footnote{ The nova
speed class is defined by $t_2$ or $t_3$,
%%% (days of 2 or 3 mag decay from the optical maximum),
for example, very fast novae ($t_2 \le 10$ days),
fast novae ($10 < t_2 \le 25$ days),
moderately fast novae ($25 < t_2 \le 80$ days),
slow novae ($80 < t_2 \le 150$ days),
and very slow novae ($150 < t_2 \le 250$ days),
as defined by \citet{pay57}, where $t_2$ or $t_3$ is the 2 or 3 mag
decay time from the $V$ maximum.} from left to right,
V1764 Her (very fast novae), KT Eri (very fast nova), V723 Cas
(very slow nova), and PU Vul (flat peak nova).
  
These novae are aligned with the decreasing order of the WD mass, i.e.,  
V1674 Her \citep[1.35 $M_\sun$; ][]{kat25v1674her}, 
KT Eri \citep[1.3 $M_\sun$; ][]{hac25kw},  V723 Cas  \citep[0.58 $\pm
0.07~M_\sun$; ][]{iij98}\citep[$\sim 0.6~M_\sun$; ][]{kat11drag}.
% from the absolute B magnitude
%, 0.59 $M_\sun$ \citet{hac04} from light curve fitting).
PU Vul \citep[$\sim 0.6~M_\sun$; ][]{kat11,kat12}. 
%and CN Cha \citep[0.58-0.6 $M_\sun$; ][]{kat23}. 
Table \ref{table_novae} summarizes the basic properties of each nova,
i.e., outburst year, distance modulus in the $V$ band,
extinction, distance, WD mass, 
binary nature, orbital period, and reference of the distance modulus
in the $V$ band.

Unlike classical novae, PU Vul and CN Cha show a long lasted flat peak
during which no massive winds are accelerated. 
In the flat peak of PU Vul, its spectra show no P Cygni profiles 
\citep[e.g.,][]{kan91b}.  

\citet{fri02} argued that V723 Cas and HR Del have a similar type of the
light curves, i.e., it starts from a slow rise which lasts several months
and then reaches a sharp maximum more than 1-2 mag above the light curve
plateau, followed by later flares. 
V723 Cas shows, after the light maximum, very similar spectral evolution 
to that exhibited by normal classical novae \citep{iij98}. 
%\citet{iij96,97}
%
Such properties in the early and intermediate phases are different 
from those in most classical novae, and 
the authors argued that they are a borderline nova for a classical nova.  
Several years later, V5558 Sgr 2007 comes into this group that shows
similar properties to V723 Cas and HR Del. 
Recently, \citet{ayd23} presented a series of spectra for Gaia22alz (Nova
Vel 2022), from the rising phase to a middle phase of the outburst.
With these data being analyzed, we regard that Gaia22alz belongs to
the same group as V723 Cas, HR Del, and V5558 Sgr.  
These similarities suggest that they originate from the same physical
mechanism.  We dub them ``metamorphic novae`` for convenience, 
based on \citet{kat11drag}'s theoretical model 
that such a light curve can be understood by a drastic envelope 
structure change from an early static envelope to a later steady state wind 
envelope through a violent relaxation process.\footnote{Violent relaxation
is a concept introduced by \citet{lyn67},
primarily in the study of elliptical galaxies. It describes a process that
occurs when a self-gravitating system is significantly perturbed from
equilibrium. During this process, the system undergoes large oscillations,
which lead to a redistribution of energy among its constituent stars or
particles.}
Thus, we categorize novae into a few classes as listed in Figure
\ref{category}, based on the WD mass and orbital separation.

%before a sharp maximum to massive wind evolution after multiple peaks 
% as depicted in Figure \ref{novae}(a).  
% (green caterpillar) (adult butterfly) (pupa) 

In the present paper, we summarize the properties of metamorphic novae, 
and analyze light curves and spectral features
of Gaia22alz to clarify when and how the winds are accelerated. 
This paper is organized as follows. 
In Section \ref{metamorphic_novae}, we examine various properties of 
V723 Cas, HR Del, V5558 Sgr, and Gaia22alz and draw common properties  
of metamorphic novae.  Our nova evolution model
for Gaia22alz is introduced in Section \ref{sec_model} and 
light curve fitting of our model with the observation is presented in
Section \ref{sec_lc}.
In Section \ref{sec_aydi}, we discuss the idea proposed by \citet{ayd23}
that the slow optical rise of Gaia22alz is owing to the irradiation
of X-ray/UV emission from the WD.  Discussion and conclusions follow in 
Sections \ref{sec_discussion} and \ref{sec_conclusion}, respectively.

%%%%%%%%%%%%%%%   Section 2 

% Fig.3
\begin{figure}
%\epsscale{1.05}
\epsscale{0.95}
%\plotone{light.4nova.f.ps}
\plotone{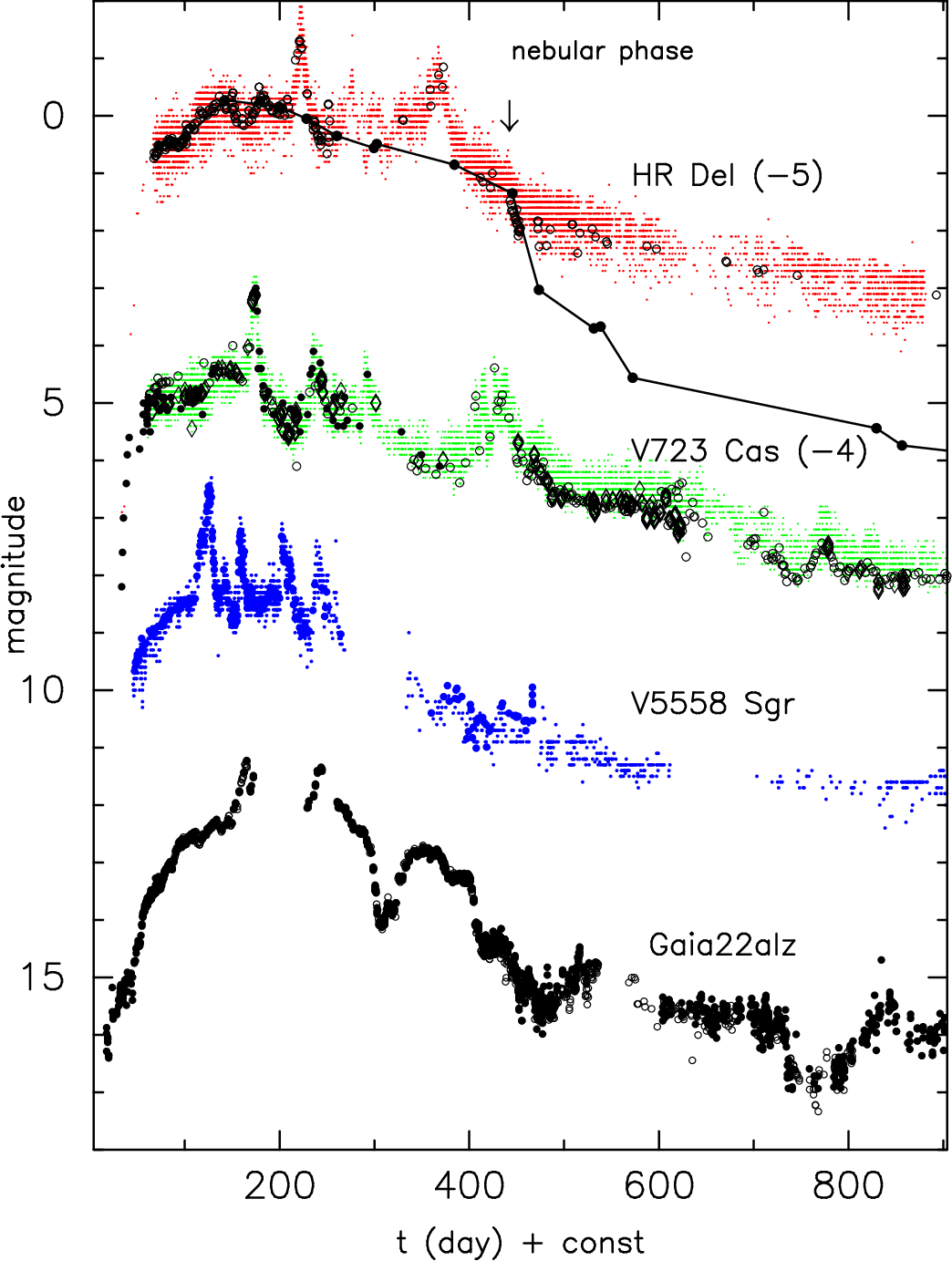}
%\plotone{light.5novae.ps}
\caption{
Comparison of $V$ light curves among the four metamorphic novae. 
HR Del: the optical ($V$ and visual) data are the same as those
in Figure 6 of \citet{kat11drag}, and  
% red dot (AAVSO), small open black circle (V) 
%(IBVS 226,238,313, 371), 
filled circle connected by black lines indicate the continuum-only 
$V$ band magnitudes taken from \citet{dre77}.
The nebular phase started on UT 1968 July 21 \citep{san74}.
These data are shifted up by 5 mag.
V723 Cas: all the magnitude data are the same as those in Figure \ref{novae}, 
%green dots (visual), 
%% two arrows for upper limit and open diamond (iauc), 
%filled black circles, open black circles,
%%%(IAUC Nos.
%%%6213, 6214, 6227, 6233, 6256, 6283, 6331, 6358, 6428),
%and open black diamonds ($V$) 
and are shifted up by 4 mag.
V5558 Sgr: blue dots (AAVSO visual) and blue filled circles
(AAVSO $V$) are left as it is, that is, not shifted. 
%shifted upward by 3 magnitude, and 
Gaia22alz (Nova Vel 2022): filled circles (ASAS-SN sky-patrol $g$) 
and open circles (AAVSO $V$) are shifted down by 0.45 mag. 
}\label{4novae}
% source:paper/26gaia22alz/light.4nova.f.wip
\end{figure}

%\clearpage

\section{Metamorphic novae} \label{metamorphic_novae}

We discuss various observational properties among HR Del, V723 Cas,
V5558 Sgr, and Gaia22alz to draw their common properties.

\subsection{Orbital periods}

Table \ref{table_novae} lists the orbital period of each nova, the source of
which is listed in each reference. 
PU Vul and CN Cha, that show a long lasted flat peak, are wide binaries. 
This means that the WD photosphere does not reach the companion orbit 
even it expands to the maximum radius \citep{kat11, kat12, kat23}. 
On the other hand,  
HR Del and V723 Cas are close binaries with orbital periods of 
$P_{\rm orb}= 0.214$ day ($=$ 5.14 hr) \citep[][and references therein] 
{fri10,kue88} and $P_{\rm orb}=0.693$ day ($=$ 16.6 hr) \citep{gor00},  
respectively. 
For V5558 Sgr, \citet{nai07} suggested that V5558 Sgr hosts a dwarf companion
from a large outburst amplitude ($\Delta V > 10$ mag).  
15 years later, \citet{schaefer22} obtained the orbital period
to be 0.1858 day ($=$ 4.46 hr). 
%%%Gaia22alz is suggested to have a dwarf companion \citep{ayd23}.  
For Gaia22alz, we adopt the orbital period of $P_{\rm orb}=1.437$ day 
from TESSExtractor \citep{ser26} as will be introduced
in Section \ref{sec_companion}. 
Thus, these four novae are close binaries. 

%For V5558 Sgr, the orbital period is  $0.185808 \pm 0.000008$ day \citep{schaefer22}. 

\subsection{WD masses}

%They estimated the WD mass in these three novae to be about 0.6  $M_\sun$, 
%having strong nova outbursts as a thermonuclear runaway event \citep{fri92,tkat02,tan11}.

The WD masses of HR Del, V723 Cas, and V5558 Sgr are estimated
by various methods. 

For HR Del, the WD mass is estimated to be
0.9  $M_\sun$ by \citet{bru82} from the orbital radial velocities, 
0.595 $M_\sun$ by \citet{kue88} from the orbital radial velocities. 
\citet{rit98} summarized them to be $M_{\rm WD}=0.67\pm 0.08 ~M_\sun$. 

For V723 Cas, \citet{iij98} estimated the WD mass to be
0.58 $\pm 0.07~M_\sun$ from the absolute B magnitude. 
\citet{hac04} obtained 0.59 $M_\sun$ from the model light curve fitting.  

For V5558 Sgr, \citet{pog10} obtained the WD mass to be 0.58-0.63 $M_\sun$
from a relation between the absolute magnitude at optical maximum
and the WD mass.

For Gaia22alz, in Section \ref{sec_model}, we estimate the WD mass
to be about 0.6 $M_\sun$ with our theoretical model light curve fitting,
because its WD mass has not been estimated.

\subsection{Resemblances in light curves}

% Fig.4
\begin{figure}
%\epsscale{1.05}
\epsscale{1.}
%\plotone{light.overlapp.abs.ps}
\plotone{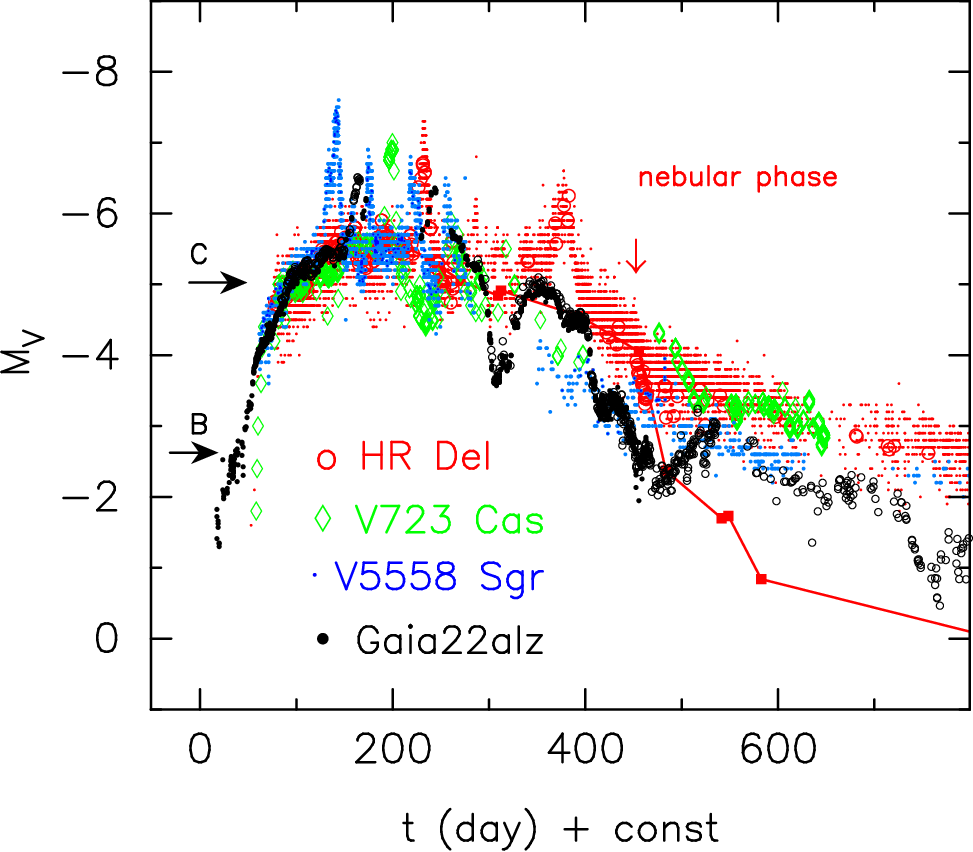}
\caption{
Same as those in Figure \ref{4novae}, but in the absolute $V$ magnitude. 
Each apparent brightness is converted to absolute brightness
with the distance modulus in the $V$ band of $(m-M)_V$ tabulated
in Table \ref{table_novae}.  The rightward arrows indicate 
epoch B and epoch C of Gaia22alz corresponding to those
in Figure \ref{othermodels}.
 }\label{novaoverlap}
% source:  nova/gaia22alz/light.overlapp.f.wip
\end{figure}

Figure \ref{4novae} depicts light curves of HR Del, V723 Cas, 
V5558 Sgr, and Gaia22alz. 
These four light curves are similar both in their shapes and timescales.
To demonstrate their similarities, we replot these four novae 
in Figure \ref{novaoverlap},  
in the absolute $V$ magnitude, $M_V$, 
converted from apparent magnitude, $m_V$, with the distance modulus 
in the $V$ band of $(m-M)_V$ listed in Table \ref{table_novae}. 
The $(m-M)_V$ of Gaia22alz  is determined in Section \ref{sec_distancemodulus}
by our model light curve fitting. 
We shift each light curve in the horizontal direction 
and overlap it with the rising phase of Gaia22alz.

Many authors have ever discussed on the resemblance among
V723 Cas, HR Del, and V5558 Sgr. 
In the multiple peaks, these novae show 
a similar amplitude of $V$ magnitude variation $\Delta V \sim 2$ 
around a similar mean $V$ magnitude of $M_V\sim -5.5$, with a 
similar variation timescale \citep[e.g., ][]{fri92,fri02,eva03,mun07,das15}.
Gaia22alz also shows the same properties, i.e, similar amplitude 
of variation around the similar mean level of $M_V\sim -5.5$,
as depicted in Figure \ref{novaoverlap}. 
 
After day 400, Gaia22alz decays faster than the other novae, 
and the lower end of the light curve is close to the continuum-only 
light curve of HR Del indicated by the filled red squares connected by 
the red line in Figure \ref{novaoverlap}.

This continuum-only $V$ light curve of HR Del is obtained by \citet{dre77}. 
In the multiple peaks of HR Del, there are no observations, 
but this continuum-only $V$ light curve agrees well 
with the $V$/visual light curves in the earlier phase than the start
of the nebular phase (from day $\sim$450).
HR Del enters the nebular phase at the downward red arrow in 
Figure \ref{novaoverlap}.
In the nebular phase, very strong emission lines such as
[\ion{O}{3}] 4959,5007\AA\  and [\ion{N}{2}] 5755\AA\  
become prominent and significantly contribute to 
the $V$ magnitude. Thus, the  $V$/visual light curves decays much 
slower than the continuum-only $V$ light curve.

The spectrum of Gaia22alz on day 497 
shows an appearance of weak [\ion{O}{3}] 5007\AA\   line indicating 
the start of the nebular phase \citep{ayd24}.  
We have no later spectra to see how the nebular lines contribute. 
We suppose that the contribution is not much to the $V$ band
in the following decline phase although it is obscured by the large
amplitude variation.  This is because the $B$, $V$, and $I_{\rm C}$ 
light curves of the AAVSO archive show a similar shape of decline
after day 400 and later.

\subsection{Resemblances in spectral evolutions}

The resemblances in the spectral evolutions among HR Del, V723 Cas,
and V5558 Sgr have been discussed by many authors: 

\noindent
{\it HR Del:} it shows several optical peaks, each of which accompanies
a discrete shell ejection \citep[e.g., ][]{san74}.  
The early spectra before the optical maximum were characterized by relatively 
narrow emission lines with a P Cygni profile, and the spectral 
class of F0 \citep{san74,raf78}. 
% However, the pre-maximum development suggests very low velocities
% near an almost stationary photosphere, unlike for the majority of
% classical novae in this stage.
\citet{fri92} wrote ``the pre-maximum development suggests
very low velocities near an almost stationary photosphere.''
%optically thick winds unlikely 
%occurred before maximum and that the photosphere is almost stationary.
% but cannot read in ADS
HR Del enter the nebular phase at the downward arrow in Figure \ref{4novae}. 
%In the nebular phase, 
After that, the spectral evolution becomes similar to normal novae
\citep{dre77,raf78}.

\noindent
{\it V723 Cas:} it shows very narrow emission/absorption lines in the spectra
at the pre-maximum flat phase. 
The spectrum gradually changes to
a pure absorption F-type supergiant before the optical maximum 
\citep{iij98}. After that, the spectrum changes to that of 
usual novae with many emission lines, some of which show 
a P Cygni profile \citep{iij98,iij06,eva03}. 

%(move figure caption)
%It entered the nebular phase (645-677 days after the discovery).
%Figure \ref{4nova} shows these two dates. 
%(Aug. 24 1995 by M.Yamamoto, IAUC No 6213).
% 2449953.5  

\noindent
{\it V5558 Sgr:} The spectral evolution before and after the optical
maximum is also very similar to that of V723 Cas 
\citep{iij07b,nai07,mun07,pog08,pog10,tan11}.
In the pre-maximum phase, the nova shows a featureless F-type spectrum 
except narrow emission lines H$\alpha$ and H$\beta$ and many weak 
absorption lines, which does not resemble to those of classical novae.
In the multi-peak episodes, fast hot winds (FWHM$=$1150 - 1500 km~s$^{-1}$)
emerge from the WD photosphere, which suggest mass ejections 
associated with the brightening \citep{mun07,kis07,tan11}.
After the multi-peak phase ended, V5558 Sgr began to decay and soon 
entered the nebular phase.  Its spectra resemble to those of V723 Cas
\citep{pog10}.

\noindent
{\it Gaia22alz:} 
\citet{ayd23} presented a series of spectrum from mid rising phase 
to the beginning of the decay phase. 
We marked epoch B and epoch C in Figure \ref{novaoverlap}, 
the exact definitions of which are given later in the beginning of 
Section \ref{sec_lc}.
The spectrum of Gaia22alz at epoch B shows no P Cygni profile,  
as is observed in the other metamorphic novae.
From epoch B to epoch C,  
the spectra are similar to those 
%\ion{He}{1} 5876 \AA~ changed from emission to absorption,  
%and weak [\ion{O}{2}] emission lines appear.  
%Similar spectral changes are reported 
in the rising phase of V5558 Sgr \citep{tan11}. 
Also, the spectrum at epoch C, characterized by strong 
H$\alpha$, H$\beta$, and weak \ion{Fe}{2} lines \citep[Figure 2 in ][]{ayd23},
is almost identical to that of V723 Cas
in the corresponding stage \citep[Figures 3 and 4 in ][]{iij98},
and similar spectral features are also reported
in the pre-maximum spectra of V5558 Sgr \citep{pog08,pog10,tan11}.

%relatively narrow Balmer and \ion{He}{1},
%with absence of P Cygni profile.  These weak lines disappear on day 68.
%
%At epoch C on the shoulder after the rising phase,  there appear 
%weak and narrow emission lines such as \ion{Fe}{2} 4584, 4629, 5169 \AA.  

%\citet{tan11} presented six pre-maximum spectra
%in the late rising phase of V5558 Sgr. 
%The Gaia22alz spectra from epoch B to epoch C agree well with the 
%gradual spectral changes of V5558 Sgr toward the optical maximum.  

% Slightly later of epoch B \citet[their day 68][]{ayd23} 
% He II 4686 \AA disappears, and weak P Cyg absorption appears 
%in H$\beta$ and H$\alpha$ lines. 

In the multi-peak phase, all the four novae show an  
indication of massive mass loss characterized by prominent P Cygni profiles 
like in normal novae, which are already reported
by the many authors cited above.  
The spectrum of Gaia22alz on day 497 \citep{ayd24} shows 
new emission lines such as [\ion{O}{3}] 5007, 
\ion{He}{2} 5412, and [\ion{N}{2}] 5755\AA\ , 
which resembles to those in the decay phase of V723 Cas 
\citep[Figure 11 in ][]{iij06}. 
%     may 30 1997 He II 5412 weak iijima
% very similar to july 11 1997 both two lines strong  
The appearance of weak [\ion{O}{3}] 5007\AA\  line indicates the start of 
the nebular phase around day 497.  
To summarize, Gaia22alz shows a spectral evolution similar to HR Del, 
V723 Cas, and V5558 Sgr.

\subsection{Comparison between metamorphic novae and flat peak novae}
\label{subsec_no_winds}

\citet{kat11drag} proposed that a metamorphic nova evolves almost 
hydrostatically in the rising phase,
and later changes to a wind evolution. 
If the transition did not occur, these novae underwent an evolution of
a flat peak nova.  In other words, a metamorphic nova follows
an evolution of a flat peak nova until the transition occurs. 
Here we see the similarities between the metamorphic novae and flat peak novae. 

First of all, both metamorphic novae and flat peak novae show 
a smooth rising phase.  The flat optical peak of PU Vul and CN Cha 
are in the range of $M_V = -5.4$ - $ -5.7$, as shown in Figure \ref{novae}(b). 
This brightness is similar to the mean level of the four metamorphic novae 
in the large amplitude multi-peak phase.

The spectrum of PU Vul resembles to the rising phase of the metamorphic novae. 
The spectrum is similar to an F supergiant without mass loss (no P Cygni profile). 
In the early flat maximum phase, it shows many narrow absorption lines 
such as Balmer and heavy elements, Fe, Si, Ti, \ion{Ca}{1} lines. 
In the later flat maximum phase, these absorptions 
gradually become shallower, and emission lines such as 
Balmer, \ion{Fe}{2} and [\ion{Fe}{2}] dominate
\citep[i.e.,][]{bel82,bel89,iij89,kan91a,kan91b}.   
% 1987
These spectral features are similar to those in the early phases 
of V5558 Sgr \citep{tan11} and Gaia22alz \citep{ayd23}. 

In CN Cha, 
% also shows a flat optical peak at $m_V \sim 8$ mag that lasted 
%three years followed by a rapid decline.  
we have only one optical spectrum \citep{lan20},
the date of which is indicated by 
the downward arrow in Figure \ref{novae}(b). 
%(JD 2458554.5:(on UT 2019 March 12,  
The spectrum shows narrow emission lines including H$\alpha$ line of 
$\sim$120 km s$^{-1}$ width with an absorption line at $\sim -20$ km s$^{-1}$, 
that we regard as the absorption by cool red-giant (RG) winds 
from the companion Mira variable. This spectrum indicates that 
%Although there is only one spectrum in the decay phase, we may conclude that
CN Cha had not experienced a massive wind evolution in the declining phase. 

%Such a spectrum feature is very different from typical classical novae 
%in which the spectrum accompany broad P-Cyg profile owing to the wind mass loss 
%0.11    1829139027066224640   PU Vul
%0.19088939      0.03931089      11.131271
%4233.942        3684.844        4840.693
%4090.013        3491.823        4629.874        10033

%in the optical decay phase.  
%  CN Cha Gaia eDR3
%  5199012973569013120
%  0.3222786       0.019822778     7.44369
%  3051.338        2876.670        3245.439
%  3062.400        2854.303        3293.817        10033

\subsection{Observational Summary of Metamorphic Novae}
\label{sec_obs.sum}

Here we summarize observational properties of metamorphic novae, 
HR Del, V723 Cas, V5558 Sgr, and Gaia22alz. 

\noindent
(1) Strong similarity in the light curves: After a smooth rising, 
  they show a large amplitude multi-peak variation of
  $\Delta V \sim 2$ around the mean brightness level of $M_V\sim -5.5$ that
  lasts a few hundred days. 

\noindent
(2) Strong similarity in the spectral evolutions: 
  In the rising phase, their spectra show no indication
  of massive wind mass loss, while in the decay phase, they
  show massive wind mass loss with P Cygni profiles similar
  to classical novae.   

\noindent
(3) Their orbital periods are less than a few days. 

\noindent
(4) Their WD masses are as low as $\sim 0.6~M_\sun$.

In short, the metamorphic novae show similar properties to the flat 
peak novae in the rising phase, whereas they show resemblance to classical 
novae in the final declining phase.

%%%%%%%%%%%%%%%  section 3

% Table.2
%\startlongtable
\begin{deluxetable*}{llllllllllll}
%\begin{deluxetable*}{llllllllllcclllll}
\tabletypesize{\scriptsize}
\tablecaption{White dwarf models of metamorphic novae in the rising phase
\label{table_model}}
%\tablewidth{0pt}
\tablehead{\colhead{Model} &
\colhead{$M_{\rm WD}$}&
\colhead{$M_{\rm env}$} &
\colhead{$X$}&
\colhead{$Y$}&
\colhead{$Z$} &
\colhead{$X_{\rm CNO}$} &
\colhead{$X_{\rm Ne}$}&
\colhead{$L_{\rm ph}$\tablenotemark{a}} &
\colhead{$L_{\rm nuc}^{\rm max}$\tablenotemark{b}} \\
%\colhead{$M_{\rm env}$ }\\
%\colhead{reference\tablenotemark{a}} \\
\colhead{ }&
\colhead{($M_\sun$) } &
\colhead{($10^{-4}~M_\sun$) }&
\colhead{ }&
\colhead{  } &
\colhead{ }&
\colhead{ }&
\colhead{ }&
\colhead{($10^4~L_\sun$) }&
\colhead{($10^7~L_\sun$) }&
}
\startdata
M06.Ne.047  & 0.6  & 0.47 & 0.55 & 0.3 & 0.02 & 0.1 &0.03  &  1.16  & 0.18 \\
M06.Ne.11   & 0.6  & 1.1 & 0.55 & 0.3  & 0.02  & 0.1 &0.03 &  1.08  & 2.8 \\
M06.Ne.25   & 0.6  & 2.5 & 0.55 & 0.3  & 0.02  & 0.1 &0.03 &  0.976 & 350 \\
M06.sol.11  & 0.6  & 1.1 & 0.7 & 0.28  & 0.02  & 0   &0    &  0.967 & 0.21 \\
%M055.Ne.047 & 0.55 & 0.48& 0.55 & 0.3  & 0.02  & 0.1 &0.03&$1.02\times 10^4$ $4.7\times 10^4$\\
M055.Ne.11  & 0.55 & 1.1 & 0.55 & 0.3  & 0.02  & 0.1 &0.03 & 0.942  & 0.71\\
\enddata
\tablenotetext{a}{Photospheric luminosity at point C ($\log T_{\rm ph}$ (K)=4.0)}
\tablenotetext{b}{Nuclear luminosity at point A ($\log T_{\rm ph}$ (K)=5.0)}
%\tablenotetext{a}{\citet{kat25v1674her}, 2: \citet{hac25kw},
%\tablenotetext{d}{Increased carbon mass-fraction by 0.1 at ignition.}
\end{deluxetable*}

\section{Model light curves in the early rising phase}
\label{sec_model}

\subsection{Theoretical Background of Metamorphic Transition}

 Nova outbursts are thermonuclear runaway phenomena of a hydrogen-rich
envelope on a mass accreting WD. 
Numerical calculations show that a hydrogen shell flash occurs 
not only on a massive WD but also on a low mass WD, i.e., on a
0.4 $M_\sun$ WD \citep{yar05} or 0.51 $M_\sun$ WD \citep{chen19}.  
However, no optically thick winds are accelerated for such a low mass WD.   
The nova undergoes a hydrostatic evolution throughout the outburst.
The nova stays at the flat optical peak for a long time,
keeping a red-giant like structure \citep{kat85,kat94h,kat09}. 

This flat peak lasts a long time, depending on the ignition mass. 
For example, if the ignition mass is as large as $M_{\rm ig} =$ a few 
times $10^{-4}~M_\sun$, the flat peak continues for $\sim$1000 yr,
because it evolves with a timescale of nuclear burning 
($\dot{M}_{\rm nuc}=$ a few times $10^{-7} ~M_\sun$ yr$^{-1}$).
Such an outburst could not be recognized as a nova outburst,  
but may be regarded to be an F supergiant star. 
On the other hand, if the ignition mass is much smaller, the duration
of the flat optical peak is as short as several years, which can be detected 
as a flat peak nova like PU Vul (8 yr optical peak) and CN Cha (3 yr),
as shown in Figure \ref{novae}(b). 

Note that, for massive WDs, strong optically thick winds emerge from
the WD photosphere and take mass quickly away from the WD envelope.
The evolution timescale is determined by the wind mass loss rate 
($\dot{M}_{\rm wind} \gtrsim 10^{-4}~M_\sun$ yr$^{-1}$). 
Therefore, the nova evolves very fast, which results in  
a fast decline in the $V$ light curve as in V1674 Her and KT Eri
\citep{kat25v1674her,hac25kw}, as shown in Figure \ref{novae}(a). 

\citet{kat09} theoretically suggested that the border line of WD mass is
$\sim 0.6~M_\sun$ which demarcates the WD mass boundary between
flat peak novae (no optically thick winds) and normal classical novae
(massive winds), depending slightly on the chemical composition
of the WD envelope. 

% Friedjung?

%The WD close to the border WD mass, Two different envelope co-exist 
%for the same envelope mass and chemical composition. 
%\citet{kat09} concluded that if a nova started its outburst from 
%the static solution, it undergoes flat peak evolution without mass loss. 
%If it started from a wind solution, it evolves as normal classical nova 
%with massive mass loss.   

\citet{kat11drag} proposed a transition mechanism 
from a flat peak evolution to a massive wind evolution. 
If PU Vul was a close binary, the expanding envelope engulfs the companion
and its orbital motion disturbs the envelope
structure, which 
%causes a large perturbation on the WD envelope to 
triggers a transition.
% to massive wind evolution.
%Multiple optical peaks suggest a relaxation process in the transition.
\citet{kat11drag} presented the light curve models of HR Del, V723 Cas,
and V5558 Sgr 
and interpreted that their large amplitude variation (multi-peak) phase 
corresponds to a violent relaxation process 
from a static evolution to a massive wind evolution.

% The transition could occur in a narrow range of the WD mass 
% around 0.6 $M_\sun$, 
% where the two different types of solutions exist simultaneously 
% for a given envelope mass. 

%These two types of solutions have very different structures. 
%Thus, once a nova outburst begins as a static expansion evolution, 
%it is hard to change to an optically thick wind structure. 

%On the other hand, in symbiotic novae, no transition is expected 
%because the WD envelope does not reach the companion orbit 
%even at the most extended stage. 

\subsection{Method of numerical calculation}

%or shoulder before final rising to maximum, 
%but occurs in the late decay phase. 
%Such a two-types solution condition is satisfied
%only in the narrow range ($\sim 0.6 ~M_\sun$) of the WD mass \citep{kat09}. 
%The condition of occurrence of optically thick winds  
%are discussed in detail by \citet{kat85,kat09}. 

We have calculated model light curves of Gaia22alz, 
in which the WD does not blow optically thick winds in the rising
phase. 
%To mimic a nova evolution 
%toward the pre-maximum phase, 
We make a sequence of static solutions with a given envelope mass \citep{kat09}.  
%that is assumed to be constant during the rising phase. 
We solve the equations of hydrostatic balance,
mass continuity, and radiative diffusion
%, and conservation of energy,
from the bottom of the hydrogen-rich envelope through the photosphere
assuming spherical symmetry.
We use the OPAL opacity \citep{igl96}.
Convective energy transport is calculated using the mixing length 
theory with the ratio of mixing length to pressure scale
height $\alpha=1.5$ \citep[see][for the effects of the $\alpha$ parameter
on the static solutions]{kat09}. 
We do not include the effects of binary motion or frictional 
energy release owing to the companion motion.

Our models in the rising phase are listed in Table \ref{table_model}. 
We chose 0.55 and 0.6 $M_\sun$ WDs. 
We do not know the chemical composition in Gaia22alz, but 
heavy element enrichment are reported in HR Del and V723 Cas 
\citep[see Table 11 in ][]{iij06}. 
We assume a chemical composition of the envelope  
to be uniform with Ne nova 2 (Ne2) in Table 2 of
\citet{hac06kb}, i.e.,
$X=0.55$, $Y=0.3$, $X_{\rm CNO}=0.1$, $X_{\rm Ne}=0.03$, and $Z=0.02$, 
where $X_{\rm CNO}$ and $X_{\rm Ne}$ are the extra CNO elements and extra Ne,
respectively.  For model M6.sol.11, we assume the solar composition, 
$X=0.70,~Y=0.28,$ and $Z=0.02$. 

In nova outbursts, the envelope mass,  i.e., the mass between 
the WD (CO or ONe) core surface and the photosphere, does not decrease much 
in the rising phase because convection widely occurs which 
mixes nuclear ash (helium) into the envelope \citep{kat25hsb}.
Gaia22alz shows no/little indication of mass loss in the rising phase.  
Thus, we assume the envelope mass is unchanged in the rising phase,
that is, the envelope mass is the same as the ignition mass.

The ignition mass is closely related to the mass accretion rate 
in the quiescent phase. 
For our 0.6 $M_\sun$ WDs,  the ignition mass is in the range of 
(1.3-2.7) $\times 10^{-4}~M_\sun$ corresponding to $\dot M_{\rm acc}=
3.0\times 10^{-8}$ -- $1.0\times 10^{-11}~M_\sun$ yr$^{-1}$ 
\citep[see Table 1 in ][]{hac26kmaxej}. \citet{chen19} obtained the 
accreted mass to be (1.21-2.59) $\times 10^{-4}~M_\sun$ for 0.6 $M_\sun$ WDs 
with $\dot M_{\rm acc}=3.0\times 10^{-8}$ --
$1.0\times 10^{-10} ~M_\sun$ yr$^{-1}$ assuming metal enrichment accretion.
\citet{yar05} obtained the accreted mass for 0.65 $ M_\sun$ WDs to be
8.35 $\times 10^{-6}$ - 5.40 $\times 10^{-4}~M_\sun$
for $\dot M_{\rm acc}=10^{-6}$ -- 
$10^{-12.3}~M_\sun$ yr$^{-1}$.

The orbital period of Gaia22alz is 1.437 day 
(Section \ref{sec_companion}).  In such a case, 
the companion may be a slightly evolved star from the zero-age main-sequence, 
and then we expect a relatively larger mass accretion rate 
in the quiescent phase such as  
$\dot M_{\rm acc} \gtrsim 3 \times 10^{-8}~M_\sun$ yr$^{-1}$. 
This value is consistent with the envelope mass of our 
model M06.Ne.11 ($1.1 \times 10^{-4} ~M_\sun$).

\citet{kat12} calculated PU Vul model with a 0.6 $M_\sun$ WD and 
the envelope mass at the early rising phase is 
$4.1 \times 10^{-5}~M_\sun$. This small value corresponds to 
a larger mass accretion rate, several $\times 10^{-7}~M_\sun$ yr$^{-1}$, 
consistent with the red giant companion.

In the present work, we chose three envelope masses, i.e., 
$M_{\rm env}=2.5\times 10^{-4}~M_\sun$, $1.1\times 10^{-4}~M_\sun$, 
and $4.8 \times 10^{-5}~M_\sun$ for our 0.6 $M_\sun$ WD.  
These values correspond to a small, middle, and large mass 
accretion rate, respectively.  For comparison, we calculate an additional
model, M06.sol.11 (a 0.6 $M_\sun$ WD and the envelop mass of 
$M_{\rm env}=1.1\times 10^{-4}~M_\sun$ with solar composition) and 
M055.Ne.11 (a 0.55 $M_\sun$ WD and the envelope mass of $M_{\rm env}=1.1
\times 10^{-5}~M_\sun$ with the envelope chemical composition of Ne2)
as listed in Table \ref{table_model}.

The time between two successive solutions is calculated 
as follows. 
The internal energy of the envelope is the summation of the 
enthalpy $\omega$ of the envelope and gravitational energy. 
\begin{equation}
E_{\rm int} = \omega - \int_{\rm envelope} {GM_{\rm WD}\over r} dm.
\label{equ_internalE}
\end{equation}
\noindent
The second term is mass-integrated gravitational potential from the 
bottom of the envelope up to the photosphere. 
The time between the two successive models can be calculated as 
\begin{equation}
dt = {\Delta E_{\rm int} \over  {L_{\rm in}- L_{\rm ph}} }=
{\Delta E_{\rm int} \over  {(f-1) \times L_{\rm ph}} },
\label{equ_time}
\end{equation}
where $L_{\rm in}$ is the injected energy flux to the envelope 
from nuclear burning region, and $L_{\rm ph}$ is the photospheric 
luminosity, i.e., the outgoing flux from the photosphere.
We set an inflow factor $f$ as $L_{\rm in}= f \times L_{\rm ph}$. 

In the evolution calculation of a nova outburst, hydrogen nuclear burning 
produces much more energy than the photospheric luminosity, 
$L_{\rm nuc} \gg L_{\rm ph}$, most of which is absorbed 
by the layer around/below the nuclear burning region. As a result, 
the outward energy flux is a few orders of magnitude smaller than 
$L_{\rm nuc}$ \citep[e.g., Figure 4 in ][]{kat22sha}. 
As time goes by, the outward flux from nuclear burning region 
becomes small and approaches $L_{\rm ph}$  
when the envelope reaches a thermal equilibrium.  
To mimic this situation, we approximate the inflow factor  
$f$ in Equation (\ref{equ_time}) to be large  $f=f_{\rm i}$ 
in the early stages ($\log T_{\rm ph}$ (K)$ \geq$ 5.0) 
and decreases  gradually with $T_{\rm ph}$ and approaches 1. 
For $f > 1$, the envelope expands even in hydrostatic balance
because thermal energy piles up in the envelope and as a result the pressure
increases.  For a larger $f$-value, the envelope expands faster.

More exactly, we assume $f_{\rm i}=1001$ for a model of the fastest rising 
case ($M_{\rm env}=2.5 \times 10^{-4}~M_\sun$).  
For a moderate rising ($M_{\rm env}=1.1 \times 10^{-4}~M_\sun$),
we assume $f_{\rm i}=41$, and $f_{\rm i}=2$ for a slowest rising 
($M_{\rm env}=4.7 \times 10^{-5}~M_\sun$).  
%$f=1.1$ at $\log T_{\rm ph}$ (K) $ \leq 4.1$ 
The resultant light curves show a similar tendency to the classical novae
of 1.25 and 1.35 $M_\sun$ WD models \citep{kat25hsb}, in which 
the light curves show a faster rise toward the optical 
maximum for a smaller mass accretion rate 
(i.e., a larger ignition mass).

%This value $f=f_{\rm i}$ is assumed to mimic the rising time 
%$\tau \propto M_{\rm ig}^{-3}$; 
%the rising phase depends on $\tau \propto \dot M_{\rm acc}^{-0.4}$ 
%in the nova outburst model of 1.25 $M_\sun$ \citep{kat25hsb} 
%and the ignition mass depends on 
%$M_{\rm ig} \propto \dot M_{\rm acc}^{-0.4/3}$ \citep{hac26kmaxej}.
%$\log T_{\rm ph}=5.7$ and 5.6

% Fig.5
\begin{figure*}
\epsscale{1.15}
%\epsscale{0.8}
%\plotone{light.gaia22alz.sol.ps}
%\plotone{light.gaia22alz.all.epsi}
\plotone{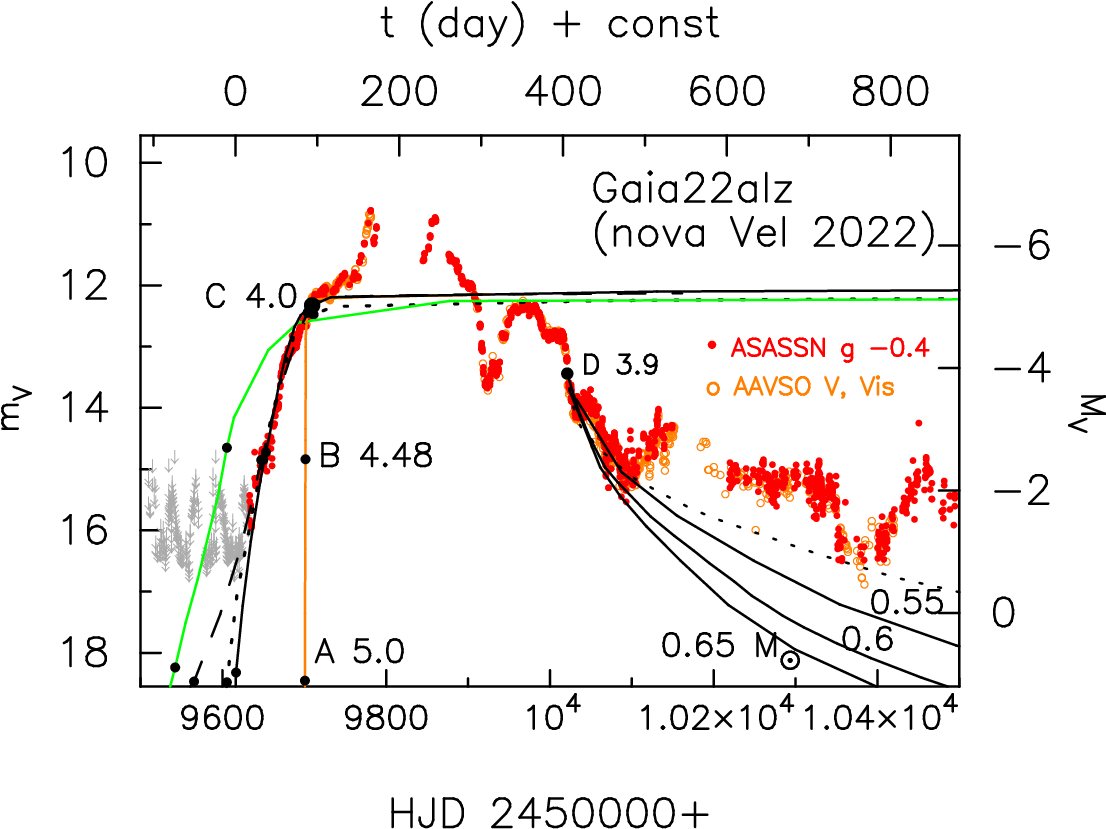}
\caption{
Optical $V$/$g$ light curves of Gaia22alz. The ASAS-SN $g$ (red circles)
and upper limits (downward gray arrows) are taken from the 
ASAS-SN sky patrol database \citep{sha14, koch17ss}.
All the ASAS-SN data are shifted up by 0.45 mag to match the AAVSO 
$V$ data (orange open circles).   The three solid model light curves
in the rising phase indicate the upper three 
models of 0.6 $M_\sun$ WDs in Table \ref{table_model}: 
M06.Ne.047 (green), M06.Ne.11 (black), and M06.Ne.25 (orange).
The lower two models in Table \ref{table_model} are 
M06.sol.11 (dashed black line) and M055.Ne.11 (dotted black line).
The rightside three solid black lines indicate the free-free emission
light curves of 0.55, 0.6, and 0.65 $M_\sun$ WDs  with 
the same chemical composition as that of M06.Ne.11 (Ne2) \citep{hac06kb} 
whereas the rightside dotted black line 
is for a 0.6 $M_\sun$ WD with the solar composition. 
These model $V$ light curves start from epoch D ($\log T_{\rm ph}
({\rm K})=3.9$).
The photospheric temperature decreases from point A (the lowest 
black dot on each model light curve: $\log T_{\rm ph}({\rm K})=5.0$)
to point B (middle black dot on each model light curve: 
$\log T_{\rm ph} ({\rm K})=4.48$), and to point C (upper black dot on
each model light curve: $\log T_{\rm ph} ({\rm K})=4.0$).
}\label{othermodels}
% source:  nova/gaia22alz/light.all.wip
\end{figure*}

\section{Light Curve Fitting of Gaia22alz }
\label{sec_lc}

%The occurrence of optically thick winds is
%checked using the condition described in \citet{kat85} and confirmed 
%that no optically thick mass-loss occurs in the rising phase.

Figure \ref{othermodels} shows the $V$ light curves in the rising phase for 
all the five models in Table \ref{table_model} (leftside model light curves). 
The absolute $M_V$ is calculated from the photospheric luminosity and 
temperature assuming blackbody emission.  The three points
on each model curve correspond to the points at 
$\log T_{\rm ph}{\rm (K)}=5.0$ (point A), 4.48 (point B), 
and 4.0 (point C). 
We overlap each point C by shifting each model 
light curve in the horizontal direction.

The light curve depends on the WD mass and chemical composition.
The three solid lines are for the same 0.6 $M_\sun$ WDs
with the same Ne2 composition but different envelope masses.  
The lower envelope mass model M06.Ne.047 (green line) evolves slowly 
because the smaller inflow energy flux corresponding to the 
smaller nuclear energy generation ($L_{\rm nuc}$, 
see Table \ref{table_model}).  
On the other hand, M06.Ne.25, a larger envelope mass model, 
shows a much faster rise. 
%This may correspond to the fastest limit of the rising phase, 
%a larger envelope mass and larger inflow energy flux results in 
%such a fast rising phase. 
Model M6.Ne.11 has an intermediate rising slope among the three
0.6 $M_\sun$ WD models, which is in good agreement with Gaia22alz.

The effects of the chemical composition can be seen 
from the comparison of M06.sol.11 (dotted black line) and M06.Ne.11
(solid black line). 
These two light curves are almost the same, indicating 
little effects of a different chemical composition. 

The dashed black line shows M055.Ne.11, a less massive WD 
but the same chemical composition and envelope mass as M06.Ne.11.  
This model light curve is also close to that of M06.Ne.11. 
Thus the difference in the WD mass ($\Delta M_{\rm WD} \sim 0.05 ~M_\sun$) 
does not substantially affect the optical light curves especially in the 
stage near point B.  

%as shown in Figure \ref{novae}. 
%Both of them belong to a symbiotic binary and the orbital period is  
%$P_{\rm orb}=4900$ days and 250 days respectively. 

\subsection{Spectral evolution of Gaia22alz (Nova Vel 2022)}
\label{sec_spectrum}
% S 4.1

\citet{ayd23} presented a series of spectra in the period from epoch B 
to day 278. 
% Figure \ref{gaia22alz.m06Ne}.  
At epoch B, the spectrum shows relatively narrow emission lines of H$\beta$,
\ion{He}{2}, and weak lines of heavy elements with no P Cygni profiles. 
In our model, the photospheric temperature at epoch B is 
$\log T_{\rm ph}$ (K)=4.48. This temperature corresponds to the 
helium ionization temperature, which naturally explains
the weak and narrow emission lines of \ion{He}{2} 4686\AA~ 
\citep[Figure 5 in ][]{tag23}. 

After epoch B, the envelope continues to expand and 
the temperature decreases to $\log T_{\rm ph}$ (K)=4.0 at epoch C. 
This temperature is too low to ionize helium and thus explains 
why the \ion{He}{2} line disappears, and instead, 
H$\beta$ emission line becomes strong between epoch B and C     
\citep{tag23}. In this way the photospheric temperature of 
our model naturally explains the spectral evolution presented 
by \citet{ayd23}. 

%The characteristic properties of the spectrum evolution of Gaia22alz,
%from day 45 to 68, is the disappearance of
%weak \ion{He}{2} 4686 \AA, \ion{C}{4}, and [\ion{O}{3}] emission lines.

Note that the absolute $V$ brightnesses $M_V$ at each point A
of all the models are almost the same.
The same thing is true at each point B (and each point C).
This is because $M_V$ is calculated from 
$L_{\rm ph}$ and $T_{\rm ph}$ with the canonical bolometric correction.  
As listed in Table \ref{table_model}, $L_{\rm ph}$ is similar 
in all the models, and thus, the resultant $M_V$ magnitude
is almost the same at each point A, B, and C.

In other words, 
%$M_V$ at each point A, B, and C is model independent. 
if we adopt another set of the model parameters (WD mass,
chemical composition, and envelope mass) to fit the observational 
light curve, 
%of the envelope mass, chemical composition and injection factor, 
we have the same $M_V$ at each point A, B, and C. 
Therefore, the above discussion on the spectral evolution with 
our model temperatures is independent of our choice of model parameters.

%\clearpage
% Fig.6
\begin{figure}
\epsscale{1.15}
%\plotone{light.gaia22alz.Ne.ps}
\plotone{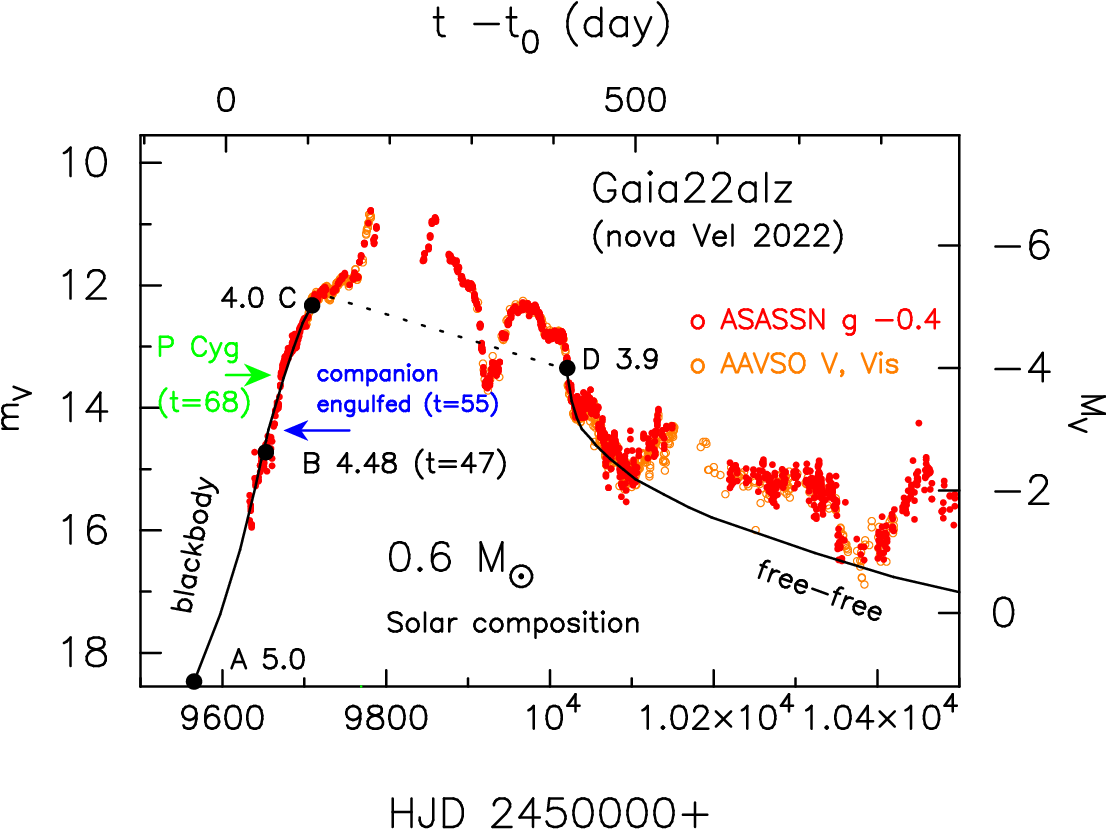}
\caption{
Same as in Figure \ref{othermodels}, but for selected models, i.e., 
the static evolution of M06.sol.11 for the rising phase, and 
steady wind evolution of a 0.6 $M_\sun$ WD (solar abundance)
for the decay phase. 
The leftward blue arrow indicates the epoch ($t=55$ day) 
when the companion is
engulfed by the expanding envelope for the binary of a 0.6 $M_\sun$ WD
and 0.2 $M_\sun$ companion with $P_{\rm orb}=$ 1.437 day. 
A P-Cygni absorption profile of H$\beta$ appears on day 68. 
A guide line (dashed line) connects the point of 
$\log T_{\rm ph}$ (K)$=3.95$
in the static evolution model and point D ($\log T_{\rm ph}$ (K)$= 3.9$)
in the steady state wind evolution model. 
We assume that a half of the initial envelope mass is lost
during the violent multi-peak phase, so it decreased from
$1.1 \times 10^{-4}~M_\sun$ at point C to $4.6 \times 10^{-5}~M_\sun$
at point D. 
}\label{gaia22alz.m06Ne}
% source:  nova/gaia22alz/light.Ne.wip
\end{figure}

% Figure \ref{gaia22alz.m06Ne} show the light curve of Gaia22alz 
% as well as our light curve model M06.Ne.11. 

\subsection{Transition from Static to Wind Evolution}

The model light curves in Figure \ref{othermodels} continue 
beyond the rightside edge of the figure ($t > 1000$ day)
%because in these models the wind mass loss is not included and 
%then the flat optical phase continues 
until the end of hydrogen nuclear burning. 
For example, the optical flat peak lasts $\sim 600$ yr in model M6.Ne.11,   
which is much longer than in PU Vul and CN Cha.

After the companion is engulfed, 
its motion and gravity disturb the envelope structure of the WD. 
The envelope may not keep its hydrostatic balance 
but could be forced to change by nonspherical effects. 
\citet{kat11drag} showed that when the companion's motion and 
gravity are taken into account, the envelope structure changes toward
that of a steady-state optically thick wind.
Thus, we may expect the occurrence of optically thick winds.

If a metamorphic transition occurs, optically thick winds can blow
most of the envelope mass in a much shorter timescale \citep{kat11drag},  
which drastically shorten the outburst duration, for example, to 
44 yr for a 0.6 $M_\sun$ WD with Ne2 chemical composition \citep{hac06kb}.

Such a transition of the envelope structure may not occur 
coherently in the whole sphere of the envelope, 
but it may begin from a small scale disturbance that triggers 
the local wind mass loss. As the companion star revolves around the WD
many times, its disturbance becomes larger and larger
and eventually triggers a transition to another structure.

\subsection{When the Wind started} \label{sec_windstart}

Figure \ref{gaia22alz.m06Ne} shows the rising phase light curve 
of model M06.sol.11. The photospheric radius increases as 
$R_{\rm ph}=0.335~R_\sun$ (epoch A, $t=-40$ day), 
3.78 $R_\sun$ (epoch B, $t=47$ day), and 34.8 $R_\sun$ 
(epoch C, $t=106$ day).
The companion star is engulfed by the expanding envelope 
when the photospheric radius reaches the orbital radius 
$R_{\rm orb}=5.0~R_\sun$. 
This occurs on $t=$ 55 day as shown by the leftward blue arrow 
in Figure \ref{gaia22alz.m06Ne}. 
Here we assume that the binary consists of a 0.6 $M_\sun$ WD and 
lobe-filling 0.2 $M_\sun$ companion 
with an orbital period of $P_{\rm orb}=$ 1.437 day 
(see Section \ref{sec_companion}).

Emergence of P Cygni profiles indicates the start of mass ejection. 
The H$\beta$ line profiles \citep{ayd23} show 
no P Cygni absorption on day 45 and 49, indicating 
no mass loss had started yet. 
The weak P Cygni profile appears on day 68, which gradually 
develops toward stage C,  
and becomes a more consistent feature with those 
of normal classical novae before the peak \citep{ayd23}.

Such spectral change is consistent with our metamorphic transition 
hypothesis that the mass loss is triggered by companion 
motion. At stage B the companion star is outside of the 
expanding WD photosphere, thus the envelope stays at a static balance and
the wind mass loss has not been trigerred yet. 
The P Cygni absorption profile deepens with time 
toward stage C. This is also consistent with our 
expectation that the wind mass loss could begin locally (non-spherically)
from a small part of the expanding envelope and develops
over the whole area of the sphere accompanying large oscillations.

\subsection{Trigger of Large Scale Transition}

After the companion star is engulfed by the expanding envelope
on day 55, it takes 51 days to reach point C. 
During this period the companion revolves the WD 35 times. 
It takes another 49 rounds from point C 
to the first prominent optical peak on $t=177$ day.   
These 35+49=84 rounds may be enough to trigger 
the structure change in a large scale.

After the first peak, the nova enters a large amplitude multi-peak phase. 
The P Cygni absorption profiles of H$\alpha$ and H$\beta$
become weaker after the peak and then become stronger again.  
Such a spectral change may be related to a local structure change. 
However, lack of $V$ light curve data (day 280 - 330) prevent further 
interpretation. 

We cannot follow such a large amplitude optical variation with our 
spherically symmetric static/steady-state numerical codes because 
the phenomena seems to be highly time-dependent, localized 
(nonspherical) structure change. 
Figure \ref{gaia22alz.m06Ne} shows the $V$ light curve of model M06.sol.11. 
With the same WD mass and the composition, the rising phase of which
is terminated shortly after epoch C,  
we calculate a wind mass loss evolution (the solid black line labeled 
free-free).  The black dashed line is just a guide line connecting
the end of a static phase (just after point C) and the start of
a steady state wind phase (point D).

% However, our aim of this work is to demonstrate that Gaia22alz 
% undergoes a static evolution with no massive (optically thick) winds in 
% the early rising phase up to a shoulder of a flat peak. 

%For massive WD like 0.7 $M_\sun$

\subsection{Decay phase}
\label{sec_decay}

%After the prominent peaks, Gaia22alz 
%gradually evolves into the decline phase. 
While the other three metamorphic novae in Figure \ref{4novae} 
show a relatively smooth decline after day 500, 
Gaia22alz seems to be still in the oscillatory phase. 
The smooth decline of the other three novae 
could be owing partly to the contribution 
of nebular line emission or partly to the end of 
large oscillatory phase.

% as in Figure \ref{gaia22alz.m06Ne}. 

%The large amplitude oscillations seem to still continue
%in the decay phase.

\citet{ayd23} regarded the decay around day 300 
%HJD 2459900 (day 300) 
as a dust blackout, without showing no observational evidence 
for dust formation.  
The AAVSO $I_{\rm C}$ light curve declined very similarly to the $V$
light curve and \citet{kar26} showed that mid-IR W1 (3.6 $\mu$m) 
and W2 (4.5 $\mu$m) bands brightness declined like optical bands,  
so we may reject dust blackout.
After the brightness recovered close to $m_v \sim 12$ 
it smoothly decays from day 412 until day 497, followed by the next 
brightening.  The spectrum 
on day 497 shows a weak [\ion{O}{3}] 5007\AA\  emission line \citep{ayd24},  
which indicates the beginning of the nebular phase.

It is difficult to determine the starting point of steady wind
evolution.
We assume that the continuum flux of Gaia22alz declines similarly to 
that of HR Del (Figure \ref{novaoverlap}), i.e. much faster than 
the light curve of nebular phase. The continuum free-free emission 
may be applicable only to a short period 
between day 400 and day 497. We choose the starting 
point of the theoretical light curves of free-free emission as 
point D. 

We plot the free-free emission light curve taken from \citet{hac06kb}.
The WD mass and the chemical composition is the same as those in 
the rising phase. The effects of binary are not included. 
In the decay phase the envelope mass is not a free parameter. 
It decreases with time owing to the wind mass loss and nuclear 
burning. Only the parameter is the starting point, which we chose 
point D. 

The envelope mass at point D of the theoretical light curve is
$4.6 \times 10^{-5}~M_\sun$. This means that 
a half of the initial envelope is lost 
during the violent multiple-peak phase, so it decreased from 
$1.1 \times 10^{-4}~M_\sun$ at point C to $4.6 \times 10^{-5}~M_\sun$
at point D.  After point D the envelope mass continuously decreases 
owing to optically-thick wind mass-loss and also to nuclear burning
\citep[see ][ for details]{kat94h}.

% Fig.7
\begin{figure}
\epsscale{1.15}
%\epsscale{0.7}
%\plotone{light.compari.rising.ps}
\plotone{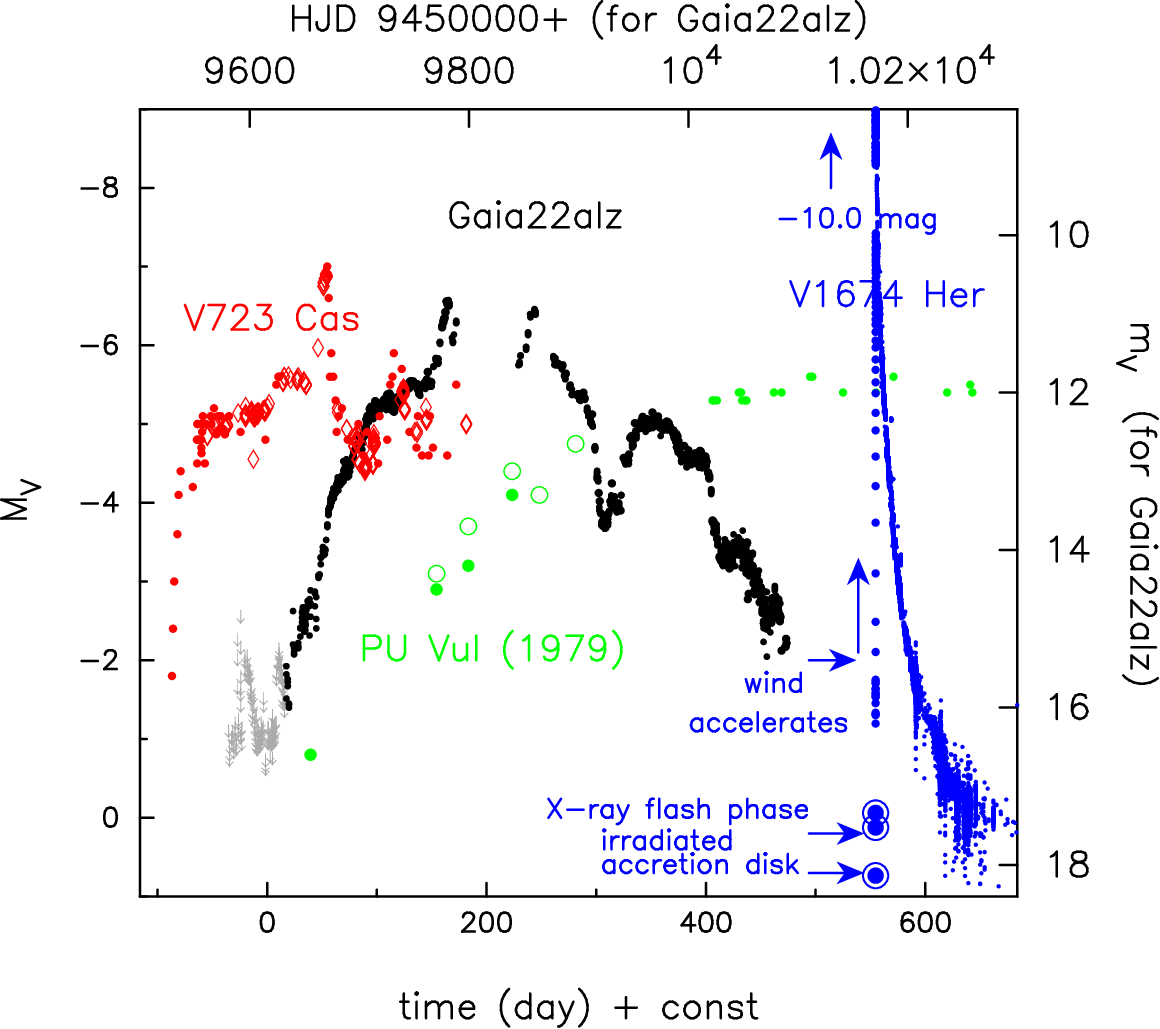}
\caption{
Comparison of the rising phases among the four novae, V723 Cas, PU Vul,
V1674 Her, and Gaia22alz. 
For V723 Cas, only the early phase is plotted by red color.  
%filled small circles are taken from IAUC Nos. 6213, 6214, 6227, 6233,
%6256, 6283, 6331, 6358, and 6428; open diamonds are from \citet{cho97}.  
%open circles (AAVSO,${\it V}$-mag), dots (AAVSO, visual).
%For PU Vul (green symbols), open circles ($m_{\rm pv}$) and  
%filled circles ($m_{\rm pg}$) are from IBVS Nos. 1608 and 1979, 
%small filled circles are from \citet{yam82}.  
%Small triangles: \citet{bel81} 
%For V1674 Her (blue symbols), large encircled filled circles are ASAS-SN $g$, 
%middle size filled circles are Evryscope $g$, small dots are AAVSO visual.  
The lower two blue rightward arrows
indicate the brightnesses of the 
irradiated disk and companion calculated for V1674 Her binary model 
with the inclination angle of $i=67 \arcdeg$ (lower arrow) 
and $45 \arcdeg$ (upper arrow) \citep{hac25kv1674her2}.  
The data of Gaia22alz are ASAS-SN sky patrol $g$ and AAVSO $V$. 
The  small gray downward arrows indicate the upper limit 
magnitudes in the ASAS-SN $g$ database.
%The gray dots are taken the ASAS-SN sky patrol database. 
%(https://asas-sn.osu.edu/sky-patrol/coordinate/
%cef95f4e-66e7-4a2b-ad65-608ca3c3ed03)}
}\label{rising}
% source: rising/light.compari.rising.wip
\end{figure}

%\clearpage

\section{E. Aydi et al.'s hypothesis of ``X-ray/UV flash detection in the visible'' }
\label{sec_aydi}

In the paper entitled ``Catching a nova X-ray/UV flash in the visible?'' 
\citet{ayd23} concluded that the ``unusual gradual rise'' of 
Gaia22alz is the evidence of reprocessed X-ray/UV photons, i.e., 
the optical flux is originated from irradiated optically thin gas 
or irradiated accretion disk. 
An X-ray flash occurs just after a thermonuclear runaway and provides
important information of this very early phase of a nova outburst. 
Thus, we examine whether this E. Aydi et al.'s statement is true or not.

\subsection{Slow rising of Gaia22alz is unusual?}

Figure \ref{rising} compares the rising phases 
of V723 Cas, Gaia22alz, PU Vul, and V1674 Her. 
Beside the very fast nova V1674 Her, 
the other three novae show a common property of no mass ejection 
(no massive winds) in the rising phase.
Among the three slow novae, PU Vul shows the slowest rising, 
followed by Gaia22alz and then V723 Cas. 
There is one more extremely slow nova ASAS J174600-232.1 
that shows a four magnitude rise in 1500 days \citep{ham15,heu15}. 
Therefore, Gaia22alz is not a rare exception of extremely slow rising 
as E. Aydi et al. noted.

% aydi et al. 2023 MNRAS unusual gradual rise  Sec 4.2 line 2
%

\subsection{Irradiation makes slow rising?}

\citet{ayd23} did not give any explanation nor even suggested possible
reasons why the optical flux gradually rises in the presence of irradiation.  
For a theoretical support of long premaximum duration, 
E. Aydi et al. referred a 0.65 $M_\sun$ WD model in \citet{hil14}. 
% in their Discussion section 4.1 
This model shows a long premaximum phase of $\sim 80$ days, 
consistent with the long and slow rising phase in Gaia22alz.  
Note that, in this model, the optical magnitude gradually 
rises from 0 to $-6$ mag toward the optical maximum 
without irradiation effects. This optical rise is roughly 
consistent with that of Gaia22alz. 
In this Y. Hillman et al.'s model, 
the X-ray magnitude $M_{\rm X}$ and UV magnitude $M_{\rm UV}$ 
\footnote{No definition is given for the X-ray/UV magnitudes.}
are fainter than the $M_V$ except the first 20 days, 
and unlikely to lift up the bright optical flux,
%except the first 20 days, 
even if all the high energy photons are
converted to optical band photons.
In other words, the model in \citet{hil14} is a counter-evidence 
to E. Aydi et al.'s explanation on the X-ray/UV irradiation.   
 
\begin{figure}
\epsscale{1.15}
%\epsscale{0.7}
%\plotone{hr.rising.ps}
\plotone{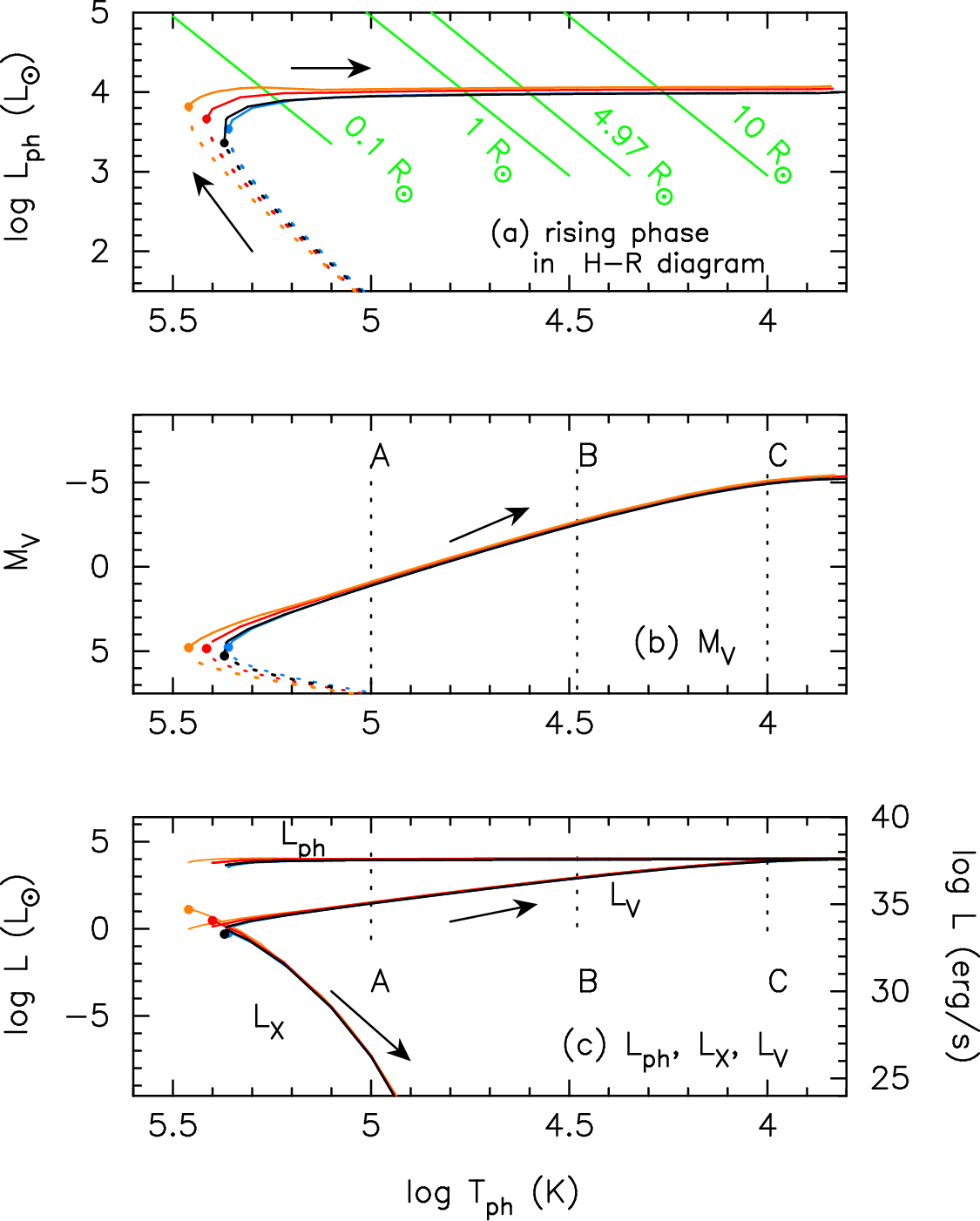}
\caption{ The H-R diagrams for (a) $L_{\rm ph}$ (bolometric luminosity), (b) $M_V$, 
and (c) $L_{\rm X}$ (X-ray: 0.3 - 1.0 keV) light curves 
of our models: M06.sol.11 (black line), M06.Ne.047 (orange line), 
M06.Ne.11 (red line), 
and M06.Ne.25 (blue line).
%  Here, the effective temperature $T_{\rm eff}$
%is the same as the photospheric temperature $T_{\rm ph}$ in our model.
%The evolutional sequence is 
The solid (dotted) line represent a hydrostatic envelope solution where
hydrogen nuclear burning is stable (unstable).  See \citet{sie75} or
\citet{nom07} for details.  The photospheric temperature reaches
maximum at the small dot.
(a) The green solid lines denote given photospheric radii. 
The line labeled 4.97 $R_\sun$ corresponds to the companion orbit of 
Gaia22alz. 
(b) Evolution of $V$ band magnitude. A, B, and C indicate the 
temperatures corresponding to point A, B, and C, respectively.
(c) Evolutions of the bolometric luminosity, supersoft X-ray luminosity
(0.3 - 1.0 keV), and $V$ band luminosity.
Only the stable part of each evolution track is plotted.
The X-ray luminosity reaches its  maximum at the 
maximum temperature (dot) throughout the stable/unstable envelope solutions.
}\label{hr.Lx}
% source: rising/hr.rising.wip
\end{figure}

Figure \ref{hr.Lx}(a) shows the evolution tracks in the H-R diagram 
of our four models (M06.sol.11, M06.Ne.047,  M06.Ne.11, and M06.Ne.25).  
As time goes by, the nova evolves upward along with the dotted 
line keeping the photospheric radius almost constant. 
Then, it turns to the right at each dot point.
Panel (b) shows the $M_V$ against $T_{\rm ph}$ for each model. 
Our light curve models begin around at point A ($\log T$ (K)=5.0). 
As the photospheric luminosity is almost constant from point A to C
the absolute $V$ magnitude $M_V$ is uniquely determined by 
the canonical $V$ response function. 
Also the X-ray flux (0.3 - 1.0 keV), calculated from the blackbody emission, 
% is the same for the same $L_{\rm ph}$ and $T_{\rm ph}$. 
hardly depends on the model as shown in panel (c). 
The X-ray flux is decreased to $\log L_X$ (erg~s$^{-1}$) $<$ 25 
shortly after point A. 
When the nova becomes observable in the $V$ magnitude 
($\log T_{\rm ph}$ (K) $\lesssim 4.5$) the X-ray flux is more than ten orders 
of magnitude smaller than the optical flux. 
Therefore, it is impossible that the X-ray irradiation of Gaia22alz 
makes its optical flux increase after it brightens up to $M_V \lesssim 1$ mag
(epoch A).

%\clearpage

% Fig.8(new)

\subsection{X-ray flash in low mass WDs}

%(YZ Ret: $1.35-1.38 ~M_\sun$, V1674 Her: $\sim 1.35~M_\sun$). 
%\citep{kat25v1674her}.  Its earliest data are 
%analyzed as the contribution of the irradiated accretion disk 
%during the X-ray flash phase \citep{hac25kv1674her2}. 

The X-ray emission can not be expected in less massive WDs because 
the photospheric temperature does not increase high enough
to emit soft X-rays. 
For example, a 0.65 $M_\sun$ WD with the mass accretion rate  
of $\dot M_{\rm acc}=1\times 10^{-8}~M_\sun$ yr$^{-1}$ shows 
the maximum photospheric temperature of
$\log T_{\rm ph}^{\rm max}$ (K) $\approx 5.56$ \citep{hil14}. 

\citet{kat22sha} presented the UV and X-ray light curves for 
their 1.0 $M_\sun$ WD model with 
$\dot M_{\rm acc}=5\times 10^{-9}~M_\sun$ yr$^{-1}$, in which 
the blackbody emission is dominated by UV (100 \AA - 300 \AA) 
photons and the X-ray luminosity is two orders of magnitude 
fainter than that because the maximum photospheric temperature 
is as low as $\log T_{\rm ph}^{\rm max}$ (K) = 5.58.  
%In this case the maximum surface temperature during the UV flash phase 
%is 370,000 K, not as high as X-ray emission.   
% 5.577 
In our 0.6 $M_\sun$ WD models, the maximum temperature 
is $\log T_{\rm ph}$ (K)= 5.370 (M06.sol.11), 
5.460 (M06.Ne.047), 5.415 (M06.Ne.11), and 
5.360 (M06.Ne.25), the positions of which are indicated by the filled
circles in Figure \ref{hr.Lx}.
Thus, the X-ray flash luminosity is about three orders of magnitude 
or more smaller than the bolometric luminosity $L_{\rm ph}$.

\subsection{Summary}
 
We found no observational evidences of irradiation effects 
in the spectra of Gaia22alz presented by \citet{ayd23}. 
The spectral evolution of Gaia22alz is consistent with the 
temperature evolution of our model. 
The slow rise in the light curve can be reproduced 
by our theoretical model without irradiation. 
Our model clearly shows that the X-ray luminosity is much smaller 
than the optical one in the observed rising phase. 
Thus, the X-ray irradiation is not required for a slow rising 
of the optical brightness as interpreted by 
\citet{ayd23}.  We clearly show that Gaia22alz is not ``unusually'' slow
in the rising phase and that its spectral evolution is not ``unusual.''
The both contradict E. Aydi et al.'s assertions.

\section{Discussion} \label{sec_discussion}

\subsection{Distance and reddening} \label{sec_distancemodulus}

% Gaia 22 alz  --- Gaia distance estimate
%
% 5307057922488654848
% 0.056318637     0.121977866     18.38514
% 6461.304        4609.913        9289.957
% 6193.986        4935.691        8020.262        10033

% Fig.9
\begin{figure}
\epsscale{1.05}
%\epsscale{0.5}
%\plotone{gaia22alz_distance_reddening.epsi}
\plotone{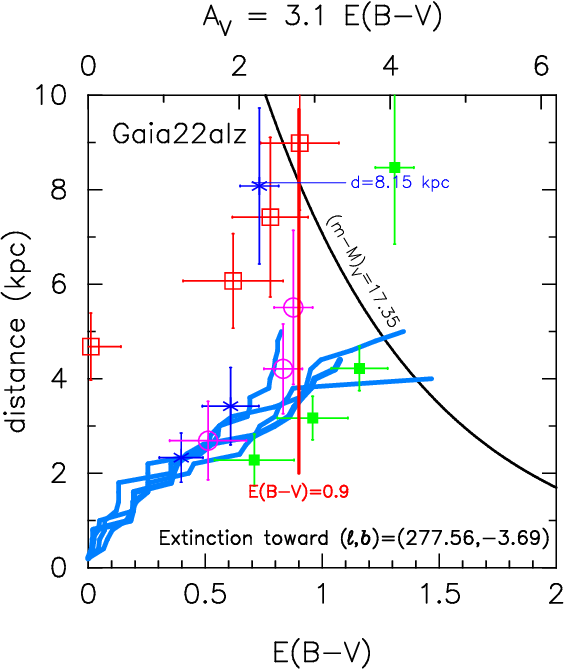}
\caption{
The distance-reddening relations toward Gaia22alz.
%Thin and thick yellow lines show the distance-reddening relations of
%\citet{gre19}.  
The black line labeled ``$(m-M)_V=17.35$'' denotes the distance-reddening
relation of $(m-M)_V= 5\log(d/{\rm 10 ~pc}) + 3.1\times E(B-V)=17.35$,
%Equation (\ref{distance_modulus_extinction}) 
%together with each $(m-M)_V$ of a nova.
where $d$ is the distance and $E(B-V)$ is the reddening.
Open red squares, filled green squares, blue asterisks, and open magenta
circles with error bars represent the relations of \citet{mar06}
in the four directions close to Gaia22alz.  
%Solid blue line indicates the distance-reddening relation of Equation
%(\ref{distance_modulus_extinction}) with each $(m-M)_V$ of a nova.
We add four distance-reddening relations (thick cyan-blue) of 
\citet{chen19hu}, which correspond to four
nearby directions toward Gaia22alz.
%  Open cyan-blue diamonds denote
%the distance-reddening relation taken from \citet{ozd16}.
%We plot four \citet{schu14}'s distance-reddening relations
%toward a nova by cyan lines with error bars.
%\citet{bai21gaia}
}\label{reddening}
% source: Izumi
\end{figure}

The distance modulus in the $V$ band is determined by direct fitting
of our model light curves with the observation.  
We have $(m-M)_V=17.35$ for model M06.Ne.11 from Figure 
\ref{othermodels}. 
For the other models in Figure \ref{othermodels}, the light curves 
are more or less similar, so this value is almost model independent. 
A larger ambiguity comes from our fitting of the model light curve with
the flat phase of the observed $V$ light curve.
We finally obtain a best fit of $(m-M)_V=17.35\pm 0.2$ by eye, 
after changing the distance modulus in the $V$ band 
from $(m-M)_V=17.0$ to 18.0 by 0.05 mag step.  
%%%Thus, this value should taken to be a rough estimate. 

%we choose the fitting point to be C, i.e., $T_{\rm ph}=10,000$ K 
%from the direct comparison of the observational and theoretical
%$V$ magnitude of model M06.Ne.11.  
%Although point C and the following flat optical peak is more or less 
%similar among the models in Figure \ref{othermodels}, 
%we don't know which the fitting level 
%the distance modulus does not change much with the choice of 
%the model parameters.  The ambiguity in the distance modulus may be 
%rather large in the choice of the fitting point than the model parameters. 

Figure \ref{reddening} shows the distance-reddening relation (black line)
for $(m-M)_V=17.35$.  %%%the distance vs. $E(B-V)$ diagram.
Figure \ref{reddening} also shows 3D distance-reddening maps
of \citet{mar06} and of \citet{chen19hu}.
%the relation taken from \citet{bai21gaia}, 
%toward the direction of Gaia22alz. 
\citet{ayd23} obtained the reddening of $E(B-V)=1.13$ toward Gaia22alz
from several diffuse interstellar bands.  They also estimated
the reddening to be $E(B-V)\sim 0.5$ from interstellar \ion{Na}{1} D
and \ion{K}{1} lines.  The value of $E(B-V)=1.13$ is close to
\citet{schlaf11f}'s Galactic 2D reddening map of $E(B-V)= 1.119\pm 0.065$
toward Gaia22alz.
It seems that the reddening does not saturate yet at the distance 
of $d \sim 8$ kpc as shown in Figure \ref{reddening}.
Therefore, we adopt the reddening of $E(B-V)=0.9$ from the cross
point between the trend of the magenta circles and red squares of
\citet{mar06} and the black line of $(m-M)_V=17.35$. 
The value of $E(B-V)=0.9$ is consistent with the reddening estimates
of $E(B-V)=0.5$ and $E(B-V)=1.13$ by \citet{ayd23}.

%0.15    5307057922488654848   Gaia22alz
%0.056318637     0.121977866     18.38514
%6461.304        4609.913        9289.957
%6193.986        4935.691        8020.262        10033
 
The cross between the black line for $(m-M)_V=17.35$ and
the vertical red line of $E(B-V)=0.9$ gives the distance of
$d=8.15$ kpc. 
The Gaia parallax ($\varpi=(0 \farcs 056318637\pm
0 \farcs 121977866)\times 10^{-3}$) of ID 5307057922488654848 star
($m_{\rm G}=$18.385) gives a geometric distance of $d=$6461 (4610-9290) pc
or photogeometric distance of $d=$6194 (4936-8020) pc
\citep{bai21gaia}.

It should be noted that the Gaia parallax gives
a reasonable/correct distance if the
parallax error is relatively small, say $<$30\% \citep{schaefer22}.
However, if the parallax is negative or its error is large,
its distance cannot be accurately constrained but depends largely
on the prior information of stars (Galactic star distribution)
in the Bayesian statistics calculation of \citet{bai21gaia}.
%%%This is the reason why a negative parallax gives a positive distance.
See \citet{schaefer18, schaefer22} for a critical explanation of
the Gaia distances.
Our distance estimate of $d=8.15$ kpc is broadly consistent with
the Gaia distance estimates.

\subsection{Companion star of Gaia22alz} \label{sec_companion}

%Whether Gaia22alz is a close binary or not is an important clue 
%to our metamorphic nova theory. 

We have determined the orbital period of Gaia22alz to be 
$P_{\rm orb}=1.437$ day using TESSExtractor \citep[A web-based interactive 
tool for light-curve visualization and period estimation using 
TESS full-frame images, ][]{ser26}. 
This orbital period is very similar to that of U Sco $P_{\rm orb}=1.23$ days
\citep{schaefer22}. U Sco is a recurrent nova that has a slightly evolved
main-sequence (or subgiant) companion of mass $\sim 1.5~M_\sun$, 
and large accretion disk 
of $\sim 2.55~R_\sun$ with $\dot{M}_{\rm acc}\sim 2.5\times 10^{-7} 
~M_\sun$ yr$^{-1}$ \citep{hac00k,hac00usco}.
From the similarity of the orbital periods, Gaia22alz possibly 
has a slightly evolved main-sequence (or subgiant) companion and 
a large size accretion disk.

The Gaia  DR3 data show
%the parallax of ID 5307057922488654848 to be 0.056318 and 
the mean $G$ magnitude $m_G=18.38$ \citep{bai21gaia}. 
%ID               parallax  
%(ID 5307057922488654848) 0.056318.. phot.g.mean.mag 18.385 
%(ID 5307057926793282432) -0.532174  phot.g.mean.mag 20.847
%
Here we assume this magnitude corresponds to the mean quiescent brightness 
of Gaia22alz.  Assuming that the $G$ magnitude is similar to
the $V$ magnitude, we obtain $m_V \approx 18.4$. 
Using our distance modulus $(m-M)_V=17.35$, we obtain $M_V=1.0$. 
Note that this $V$ magnitude is the summation of the brightness of the 
companion star and accretion disk.
This magnitude is much brighter than typical classical novae 
\citep[see, e.g., Figure 1 of ][]{war87}.

HR Del ($P_{\rm orb}=0.214$ day) has not yet reached
the quiescent brightness, but it is now close to $M_V\sim 2.3$
\citep[e.g.,][]{sel19}. Gaia22alz is much brighter than this value. 
%%% quiescent phase   M_V=0.6$  %%%%
The quiescent phase brightness of U Sco is not well determined 
%because $M_V \sim 18.6$ \citep{hac00usco} 
because its distance is not well constrained (6-11 kpc).
KT Eri in Table \ref{table_novae} is a classical nova that hosts
a massive WD \citep[$1.3~M_\sun$,][]{hac25kw} and a subgiant companion.   
It has a large accretion disk with a mass accretion rate of 
$\dot{M}_{\rm acc}\sim 1\times 10^{-9} ~M_\sun$ yr$^{-1}$
\citep{hac25kw}.
%Its orbital period is $P_{\rm orb}=2.6$ day, much longer than those of 
%typical classical novae, so it has a very large size accretion disk. 
Its quiescent brightness is $V\sim 15.1$ mag, which is converted 
to $M_V=0.5$ with the distance modulus in the $V$ band of
$(m-M)_V=14.6 \pm 0.2$ \citep{hac25kw}. 
Thus, the quiescent brightness of Gaia22alz is much brighter than 
HR Del or typical classical nova, but fainter than KT Eri.
This also suggest that Gaia22alz has a large size accretion disk.
Note again that the quiescent brightness is the summation of 
the contribution of disk and companion. When the disk is large 
the quiescent brightness is dominated by the disk contribution 
and then its color is blue. This 
is consistent with the blue color of Gaia22alz progenitor 
\citep{ayd23}.

\citet{ayd23} made a spectral energy distribution (SED) 
of the progenitor system in the wavelength region between 
3500 \AA~ to 10,000 \AA. 
They fit the SED with one component model of a Rayleigh-Jeans tail 
of a hot and very bright blackbody 
($T \sim 10^5 $ K and $L \sim 4740~L_\sun$). 
From the good fitting between 3500 \AA~ to 10,000 \AA~ with no residuals,
the authors ruled out a yellow/red giant donor star. 

However, \citet{ayd23} could not reconcile such a hot and bright
component with the blue color, large extinction, and relatively 
long distance.  
This is partly because the authors misunderstood the observed color
to be for the companion, instead of the accretion disk. 
In addition, they fit the SED with one component (blackbody) model. 
When the SED is dominated by the disk emission, 
it is difficult to find a cool component by their
one component model with a narrow range of SED ($\lambda >$ 3500 \AA). 
If one adopt two or three component model \citep[e.g.,][]{pop96}, 
the results would be very different. 

%that cannot be approximated by a single blackbody temperature. 

\subsection{Strope et al.'s classification}

% 4 F-class 
% DO Aql(0.16776)  mid phase is missing
% V849 Oph(0.17276) variable flat phase  -> metamorphic
% BT Mon(0.33381, Tess 0.328 d) , 
% V2295 Oph(none  Tess 11.445d,9.184d, total: 10.415 day)  flat peak is 10 days too short  

% J class  (schaefer2025) 
% V443 Sct(none)  V1186 Sco (0.29207) 
% V373 Sct(0.81910)  V868 Cen  (none) V2540 Oph(0.28478)
%  PW Vul(0.12858) RR Pic (0.24502) V356 Aql (0.42651)
%  V1039 Cen(0.247)   V1819 Cyg (none, tess?)   V4745 Sgr(0.20782)
% DK Lac (none)  HR Del   V723 Cas
%They listed three F-class nova involving DO Aql, but not PU Vul and 
%CN Cha, the latter appeared after 2010. 

%J-class shows ``decline displays substantial variability 
%often as short-duration brightening''. 

\citet{str10} classified 94 nova light curves into seven classes
by their characteristic shapes. 
Our metamorphic novae are closely related to J-class in their classification. 
They listed 14 J-class novae that includes V723 Cas and HR Del 
as typical examples. 
Most of them are known as close binaries \citep{sch25}. 
They could be candidates for metamorphic novae, 
but we need to see if they pass the requirements (1)-(4) 
for metamorphic novae in Section \ref{sec_obs.sum}. 

%PU Vul is discussed but excluded from their classification 
%by the reason of symbiotic nova. 
 
%V5558 Sgr is not listed. Gaia22alz is not included, but it outbursted 
%after the paper is published.
%In many novae seems to have no data in the riging phase
%it is not clear that all of these novae are the candidate of metamorphic novae. 

The difference between the classification by \citet{str10} and ours 
%in Figure \ref{category} 
are simple.  R. Strope et al.  classified novae by its light curve shape. 
%, independently to the binary nature, 
%such as WD mass, orbital period, chemical composition. 
%Such a simple classification is useful to shed light a common physics 
%independent of binary parameters. 
%
Our classification in Figure \ref{category}  
is based on the physical mechanism that drives the metamorphic transition,
in other words, the nova starts its outburst from a static envelope 
and then metamorphoses to an optically thick wind envelope after large
perturbations in the envelope.
 
The optically thick winds are accelerated by radiation pressure gradient 
\citep{fri66} which is strong in massive WDs \citep{kat94h}. 
In less massive WDs the optically thick wind does not occur, so the 
nova undergoes a static evolution, resulting in a long-lasted flat peak nova. 
In the metamorphic nova, the optically thick winds occurs 
after the envelope substantially expands beyond the binary orbit.  
In a narrow range of the WD masses,
the nova outburst starts as a flat peak nova 
but changes its structure under the disturbance by binary motion 
\citep{kat11}.  Then the nova evolves by wind mass loss.   

If this hypothesis is correct, all the metamorphic transition occurs 
on a low mass WD in a close binary. 
It is not clear if all the J-class novae in R. Strope et al.'s list 
are metamorphic novae.  But all the metamorphic novae that satisfy our 
requirements (1)-(4) belong to the J-class.  
If all the J-class novae were metamorphic novae, we could clarify
the physical mechanism of the J-class novae.

\section{Conclusions}
\label{sec_conclusion}
 
We studied characteristic properties of Gaia22alz as well as 
HR Del, V723 Cas, and V5558 Sgr along the metamorphic nova theory 
proposed by \citet{kat11drag}. 
% in which the nova starts static evolution 
%and change to steady-state evolution after the large structure change 
Our numerical codes cannot directly follow the large 
structure change of a nova, so our numerical models treat mainly the rising 
phase of a nova before the metamorphic transition occurs. 
Our results are summarized as follows: \\

\begin{enumerate}

\item 
The very slow novae HR Del, V723 Cas, V5558 Sgr, and Gaia22alz show 
remarkable similarities in their light curve shapes and spectral evolutions. 
We name these novae metamorphic novae.  

\item
The metamorphic novae show no indication of winds 
such as P Cygni profiles in the early rising phase. 
The ensuing large amplitude multiple peak episode lasts a few hundred 
days with an amplitude of $\Delta V \sim 2$ around the mean absolute 
magnitude of $M_V \sim -5.5$. The spectra show P Cygni profiles. 
After that, the optical light curve decays, and the spectral features 
become similar to those of typical classical novae. 

\item 
We present a theoretical light curve model in the rising phase
assuming that the nova envelope expands hydrostatically. 
Our light curve model shows the temperature decrease in the rising phase, 
which naturally explains the spectral features observed by \citet{ayd23}.

\item 
The weak P Cygni profiles appeared in the mid-rising phase, the epoch of 
which is consistent with the epoch when the companion star is engulfed
by the nova envelope. 
We propose that the orbital motion of the companion in the nova
envelope violently disturbs the hydrostatic structure of nova envelope 
to trigger wind mass loss.

% However, the light curve is hardly affected by the 
%presence of the weak wind.

\item
In the decline phase, we fit the observed light curve with 
a steady-state wind evolution. 
With our model light curve fittings both in the rising/decline phases,
we determine the distance modulus
in the $V$ band to be $(m-M)_V=17.35\pm 0.2$.
The distance to Gaia22alz is estimated to be
$d=8.15\pm 1$ kpc for $A_V=3.1 E(B-V)=3.1\times 0.9=2.8$.

\item 
Gaia22alz is a close binary with 1.437 day orbital period.
This supports our metamorphic nova hypothesis that the violent relaxation
from static to optically thick wind configuration is triggered by 
the companion's orbital motion in the WD envelope. 

\item We examine \citet{ayd23}'s statement that the unusual, gradual rise
in Gaia22alz could be attributed to the reprocessed X-ray/UV flash photons.
We found no indication of X-ray/UV irradiations in the optical light curves
as well as  in their spectra. Comparing with the other very slow novae,
Gaia22alz does not show an exceptionally slow rise as
E. Aydi et al. noted.

\end{enumerate}

\begin{acknowledgments}
We are grateful to the anonymous referee for useful comments
that improved the manuscript.
We acknowledge with thanks the variable star observations
from the AAVSO International Database contributed by
observers worldwide and used in this research.
%and the Variable Star Observers League of Japan (VSOLJ) for V597 Pup. 
\end{acknowledgments}

%\clearpage

% universal decline law ($L_\nu \propto t^{-1.75}$) 

%\vspace{5mm}
\facilities{AAVSO, ASAS-SN}

%%% , Gaia
%%%, ASSAS-SN, Evryscope, Gaia
%%%%  Appendix A

%\appendix

% todayreference 

\end{document}